\documentclass[aps,pra,showpacs,twocolumn,amsmath,amssymb]{revtex4-1}
\usepackage{graphicx}
\usepackage{amsmath}
\usepackage{color}
\usepackage{dcolumn}
\usepackage{bm}
\usepackage{booktabs}
\usepackage{subfigure}

\newcommand{\mHeB}{$^4$He$^*$}
\newcommand{\mHeF}{$^3$He$^*$}
\newcommand{\mHe}{He$^*$}
\newcommand{\HeB}{$^4$He}

\newcommand{\Rb}{$^{87}$Rb}

\newcommand{\etal}{\textit{et al.}}
\newcommand{\quartet}{$^4\Sigma^+$}
\newcommand{\doublet}{$^2\Sigma^+$}
\newcommand{\sss}{$F=2$, $m_F=2$}
\newcommand{\mjp}{$m_J$=1}

\begin{document}

\title{Ultracold mixtures of metastable He and Rb: scattering lengths from \textit{ab initio} calculations and thermalization measurements}

\author{S.\,Knoop$^1$}\email[]{s.knoop@vu.nl}
\author{P.\,S.\,\.{Z}uchowski$^2$}{\email[]{pzuch@fizyka.umk.pl}
\author{D.\,K\c{e}dziera$^3$}
\author{{\L}.\,Mentel$^4$}
\author{M.\,Puchalski$^5$}
\author{H.\,P.\,Mishra$^1$}
\author{A.\,S.\,Flores$^1$}
\author{W.\,Vassen$^1$}

\affiliation{$^1$LaserLaB, Department of Physics and Astronomy, VU University, De Boelelaan 1081, 1081 HV Amsterdam, The Netherlands \\
$^2$Institute of Physics, Faculty of Physics, Astronomy and Informatics, Nicolaus Copernicus University, Grudziadzka 5, 87-100 Torun, Poland \\
$^3$Department of Chemistry, Nicolaus Copernicus University, 7 Gagarin Street, 87-100 Torun, Poland \\
$^4$Section of Theoretical Chemistry, Department of Chemistry, VU University, De Boelelaan 1083, 1081 HV Amsterdam, The Netherlands \\
$^5$Faculty of Chemistry, Adam Mickiewicz University, Umultowska 89b, 61-614 Pozna{\'n}, Poland}
\date{\today}

\begin{abstract}
We have investigated the ultracold interspecies scattering properties of metastable triplet He and Rb. We performed state-of-the-art \textit{ab initio} calculations of the relevant interaction potential, and measured the interspecies elastic cross section for an ultracold mixture of metastable triplet $^4$He and \Rb~in a quadrupole magnetic trap at a temperature of 0.5~mK. Our combined theoretical and experimental study gives an interspecies scattering length $a_{4+87}=+17^{+1}_{-4}$~$a_0$, which prior to this work was unknown. More general, our work shows the possibility of obtaining accurate scattering lengths using \textit{ab initio} calculations for a system containing a heavy, many-electron atom, such as Rb.
\end{abstract}

\pacs{31.15.A-, 34.20.Cf, 34.50.Cx, 67.85.-d}

\maketitle

\section{Introduction}\label{Introduction}

Ultracold mixtures of different atomic species offer a number of advantages over single species experiments. For instance, these mixture are the starting point to obtain a dense sample of ultracold heteronuclear molecules, which (in contrast to homonuclear molecules) can have long-range and anisotropic interactions, resulting in rich new physics with many novel properties \cite{carr2009cau}. Ultracold mixtures can also feature very interesting few- and many-body phenomena, for which the mass ratio between the two atomic species can play a crucial role (see e.\,g.\,\cite{levinsen2009ads,*castin2010fbe,*mathy2011tma,*blume2012ufb,*efremov2013tbb,*zinner2013eso,*hu2014oog}). A prominent example is the observation of more than two successive Efimov loss features to test the scaling laws of the Efimov trimer spectrum, which experimentally requires an extremely large mass ratio, and for which first results have been obtained in ultracold mixtures of $^6$Li+$^{133}$Cs \cite{pires2014ooe,*tung2014oog}. 

Most experiments on ultracold mixtures involve two alkali-metal species, while recently also mixtures of alkali-metal and alkaline-earth(-like)
atoms became available \cite{nemitz2009poh,*hara2011qdm,*hansen2011qdm,*pasquiou2013qdm}. Here we are considering a different type of mixture, namely of an alkali-metal atom and helium, in the metastable 2~$^3$S$_1$ triplet state (denoted as \mHe, radiative lifetime of about 8000~s), for which quantum degeneracy has been realized for both fermionic \mHeF~and bosonic \mHeB~isotopes~\cite{vassen2012cat}. The application of \mHe~in ultracold mixtures increases the range of possible mass ratios by a factor of two compared to the commonly used $^6$Li. 

The feasibility of an ultracold or quantum degenerate mixture depends strongly on the collisional properties and stability, which in turn is given by the intraspecies and interspecies interaction potentials. Scattering between \mHe~(total electron spin $s$=1) and an alkali-metal atom ($s$=1/2) in the electronic ground state is described by a doublet \doublet~and a quartet \quartet~molecular potential. Here we focus on the \quartet~potential, for which Penning ionization is suppressed due to spin-conservation \cite{byron2010sop} and which fully describes a mixture in which both atoms are either in the lower or upper spin-stretched states. These spin-mixtures are most favorable for sympathetic and evaporative cooling towards quantum degenerate mixtures. Precise knowledge of those potentials is completely lacking, due to the absence of spectroscopic data. Therefore one has to rely on \textit{ab initio} calculations, for which however the predicted power in terms of accurate scattering lengths is generally considered to be limited, except for few-electron systems like \mHe+\mHe~\cite{przybytek2005bft}.

In this article we present state-of-the-art \textit{ab initio} quantum chemistry calculations of the \quartet~potential and the quartet scattering lengths for the \mHe+Rb system. In parallel, we have experimentally determined the quartet scattering length for \mHeB+\Rb~by measuring the interspecies elastic cross section for an ultracold mixture in a quadrupole magnetic trap. Our combined theoretical and experimental work gives tight bounds on the interspecies scattering lengths, which provides crucial knowledge for the realization of quantum degenerate \mHe+Rb mixtures. 

This paper is organized as follows. In Sec.~\ref{theory} we present the \textit{ab initio} calculations. In Sec.~\ref{experiment} we describe the experiment, including a theoretical description of interspecies thermalization measurements in a quadrupole magnetic trap (Sec.~\ref{thermalizationtheory}), the experimental setup (Sec.~\ref{experimentalsetup}), the two-species magneto-optical trap (Sec.~\ref{magneto-optical trap}) and quadrupole magnetic trap (Sec.~\ref{quadrupole magnetic trap}),  and the determination of the scattering length (Sec.~\ref{thermalizationmeasurements}). In Sec.~\ref{resultsconclusion} we compare the theoretical and experimental results and conclude. Finally, in Sec.~\ref{outlook} we give some future prospects.

\section{Ab initio calculations}\label{theory}

The calculations of the \quartet~potential have been performed using the all-electron restricted open-shell coupled cluster singles and doubles with noniterative triples corrections [CCSD(T)] method  \cite{knowles1993cct}, implemented in the MOLPRO package \cite{MOLPRO_brief}, and using the Douglas-Kroll-Hess Hamiltonian to take into account relativistic effects \cite{reiher2004edo}. These calculations are challenging for a system like He$^*$Rb, since the molecular states are submerged in the continuum of ionized states of HeRb$^+$, which might in principle lead to a variational collapse to lower lying states already during the Hartree-Fock (HF) optimization \cite{hapka2013fpi}. To circumvent this, we have constructed starting orbitals from appropriate orbitals of the isolated atoms and during the optimization we have kept the occupancies of orbitals fixed. 

For He$^*$Rb the available standard electronic gaussian basis sets are not appropriate, in particular because the basis sets for He are optimized to recover the ground-state energy. Therefore we have optimized our own basis set, suitable for \mHe. For Rb we have used the ANO-RCC basis set \cite{roos2003ran}, to which we have added one $g$- and two $h$-type orbitals optimized to the atomic energies. To better account for the dispersion interaction we have augmented both basis sets using two sets of even-tempered functions per function type (generated with the MOLPRO package). The convergence of the counterpoise-corrected interaction energies \cite{boys1970tco} is carefully analyzed both in terms of the number of augmented functions added, as well as the highest angular momentum function in the basis set. By removing the $h$ basis functions we have found that the interaction energy changes by less than 1~cm$^{-1}$.

The coupled-clusters equations are divergent for internuclear distances smaller than $r=8$~$a_0$, for which the interaction energy is approximately $-200$~cm$^{-1}$ and the inner turning point is not yet reached.  Still, we were able to converge the HF reference state down to $r=5.5$~$a_0$, from which we can exclude the possibility of crossings of the potential energy curve with other states. To extrapolate the potential towards shorter distances we have made use of the fact that the contribution of correlation energy to the interaction potential, which is by far dominated by the dispersion energy, varies exponentially near the inner turning point \cite{tang1984ais} and added the extrapolated values to the HF interaction energy (see Appendix \ref{app1}).

\begin{figure}
\includegraphics[width=8.5cm]{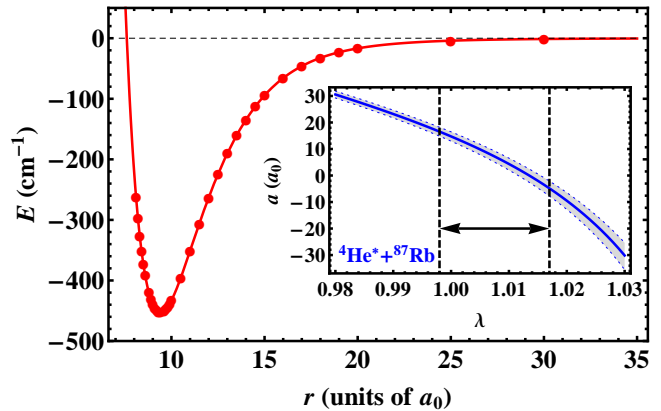}
\caption{(Color online)
\label{quartetpotential} Results of the \textit{ab initio} calculations on the \quartet~potential of He$^*$Rb (red circles) and the MLR fit (red solid line). The inset shows the quartet scattering length for \mHeB+\Rb~as function of the scaling parameter $\lambda$ (see text), where the shaded area (bounded by the blue dotted lines) represents the uncertainties in the long-range coefficients. The dashed vertical lines and the arrow represent the uncertainty in the \textit{ab initio} calculations, corresponding to a range of $\lambda$=0.998-1.017.}
\end{figure} 

The long-range van der Waals coefficients have been recently calculated by Zhang \etal~\cite{zhang2012lri}, however, with an uncertainty of 1-5\% in $C_6$ and 1-10\% in $C_8$ and $C_{10}$, which is too large for our purpose. We therefore have calculated the dipolar- and quadrupole dynamic polarizabilities of \mHe~at imaginary frequencies and integrated them with recently tabularized dynamic polarizabilities of Rb. With the present calculations the error in the $C_6$ coefficient for Rb$_2$ system is estimated as 0.5\%, while the $C_6$ and $C_8$ coefficients of He$^*_2$ reproduce the reference data \cite{yan1998lri} to better than 0.1\%. For He$^*$Rb we obtain $C_6=8.47(2)\times 10^{8}$~cm$^{-1}$$a_0^6$, with an uncertainty of 0.25\%. Using the single-pole approximation to the quadrupole dynamic polarizability derived by Porsev \etal~\cite{porsev2014rmb}, we have obtained also $C_8=8.01(4)\times 10^{8}$~cm$^{-1}$$a_0^8$, which has an accuracy of $0.56\%$. Finally we include the $C_{10}$ coefficient from Zhang \etal~\cite{zhang2012lri} to the long-range part of our potential. 

\begin{table}
\caption{Parameter values of the MLR potential.}
\begin{ruledtabular}
\begin{tabular}{cccc}
parameter	& value  																			& parameter	& value \\ \hline
$D_e$  		& 452.71~cm$^{-1}$  													& $\phi_0$ 	& -1.8284 \\
$r_e$  	  & 9.4079~$a_0$  															& $\phi_1$  & 0.48678 \\ 
$C_6$ 		& 8.4673$\times 10^{8}$~cm$^{-1}$$a_0^6$ 			& $\phi_2$  & -0.065081 \\
$C_8$ 		& 8.0108$\times 10^{10}$~cm$^{-1}$$a_0^8$ 		& $\phi_3$  & -0.30087 \\
$C_{10}$ 	& 9.4242$\times 10^{12}$~cm$^{-1}$$a_0^{10}$	& $\phi_4$  & -1.5195 
\end{tabular}
\end{ruledtabular}
\label{tab:potential}
\end{table}

The potential energy curve obtained from the \textit{ab initio} calculations is shown in Fig.~\ref{quartetpotential}. The data points are fitted with the Morse/Long-Range (MLR) potential proposed by LeRoy \etal~\cite{leroy2006aaa}, which has the form:
\begin{equation}
V(r)=D_e\left[ 1- \frac{u_{\rm LR}(r)}{u_{\rm LR}(r_e)} \exp(-\phi(r) y_p(r))\right]^2-D_e,
\end{equation}
where $D_e$ is the well depth of the potential, $r_e$ the equilibrium distance, $u_{\rm LR}(r)=C_6r^{-6}+C_8r^{-8}+C_{10}r^{-10}$,  $y_k(r)=\left(r^k-r_e^k\right)/\left(r^k+r_e^k\right)$ and $\phi(r)=\left[1-y_p(r)\right] \sum_{j=0}^4 \phi_j y_q(r) + y_p(r) \phi_{\infty} $, where $p=5$ and $q=4$. The free parameters in the potential, determined by fitting, are $\phi_j$ ($j=0,\dots,4$), while $D_e$, $r_e$ and $\phi_\infty=\log[2 D_e/u_{\rm LR}(r_e)]$ are directly obtained from the \textit{ab initio} calculations. The resulting parameter values of the MLR potential are given in Table~\ref{tab:potential}. Note that the statistical error introduced by the analytical fit is much smaller than the systematic uncertainty in the \textit{ab initio} calculations. The MLR potential is particularly convenient for the analysis of the scattering length in the case that the long-range part of the potential is known very accurately, but the short-range potential has a larger uncertainty. Hence we can conveniently parametrize the scattering length by introducing a $\lambda$ scaling parameter such that $D_e \to \lambda D_e$. 

To predict quantitatively the scattering length it is crucial to explore possible errors in the \textit{ab initio} calculations. We therefore have used higher-order coupled cluster methods, using the MRCC code \cite{kallay2001hei}, to estimate the uncertainty in the potential energy curve. We have found that the systematic error that can be attributed to the incompleteness of CCSD(T) is approximately +4.3~cm$^{-1}$ (see Appendix \ref{app2}). We have also compared the potential depths for the homonuclear He$^*$ and Rb dimers obtained with the same method used in this work, and the reference $D_e$ parameters for He$^*_2$ \cite{przybytek2005bft} and Rb$_2$ \cite{strauss2010hra}. The fractions $\delta({\rm X}_2)=\left[D^{\rm ref}_e({\rm X}_2)-D^{\rm calc}_e({\rm X}_2)\right]/D^{\rm ref}_e({\rm X}_2)$ are respectively 5.4$\times 10^{-3}$ and 4.1$\times 10^{-2}$, hence for the heteronuclear system we can estimate the error as $\sqrt{\delta({\rm He}^*_2) \delta({\rm Rb}_2)}=0.015$, which translates into +6.8~cm$^{-1}$. Finally, the long-range CCSD(T) potential curve can be tested by comparing it with the $V_{\rm LR}(r)=-u_{\rm LR}(r)$ expansion of the interaction energy. For distances between 25~$a_0$ and 40~$a_0$, where the potential energy is dominated by the interaction of multipoles, $V_{\rm LR}(r)$ is systematically larger by about 1.2\% than the CCSD(T) potential, which corresponds to a difference in $D_e$ of +5.4~cm$^{-1}$. All estimations on a possible error in $D_e$ give a systematically positive shift. Therefore we  conservatively assume the error bound between $-1$~cm$^{-1}$ (uncertainty of basis set) and +7.8~cm$^{-1}$ (uncertainty of basis set and most conservative estimate of the CCSD(T) uncertainty), which translates to a scaling parameter range of $\lambda=0.998-1.017$. 

From the potential energy curve we calculate the scattering lengths for all the four isotope combinations, for which the results are given in Table \ref{scatteringlengthtable}. The inset of Fig.~\ref{quartetpotential} shows the scattering length for \mHeB+\Rb~as function of $\lambda$. We find that the scattering lengths for all isotope combinations are small, i.\,e\, $|a|\leq 21$~$a_0$. The small difference between $^{85}$Rb and \Rb~for a given \mHe~isotope is due to the small difference in reduced mass. In contrast, the small difference between \mHeB+Rb and \mHeF+Rb, for which the reduced mass is very different, is completely accidental. For instance, the \mHeB+Rb~potential supports 15 bound states, compared to 13 for the \mHeF+Rb~potential.

\begin{table}
\caption{The quartet scattering lengths of He$^*$+Rb in units of Bohr radius $a_0$, obtained from the \textit{ab initio} calculations (theory) and thermalization measurements (experiment). For the theory we give the values connected to $\lambda$=1 and the bounds corresponding to $\lambda$=[1.017;0.998]. The experimentally obtained interspecies elastic cross section give rise to two possible values of the scattering length (see Sec.~\ref{thermalizationmeasurements}), where the error bars corresponds to one standard deviation.}
\begin{ruledtabular}
\begin{tabular}{c|cccc}
isotopes						& 3+85				& 3+87				& 4+85				& 4+87 		\\
\hline																	       
theory							& $+6$				& $+5$				& $+18$				& $+16$	\\		
										& $[-19;+8]$	& $[-21;+7]$	& $[-5;+20]$	& $[-8;+18]$	\\	
\hline									  	       
experiment 					& 						& 						& 						& $-29_{-5}^{+5}$ or $+17_{-4}^{+4}$               
\end{tabular}
\end{ruledtabular}
\label{scatteringlengthtable}
\end{table}

\section{Experiment}\label{experiment}

In the following we discuss the thermalization measurements, including the experimental setup and our strategy to obtain an ultracold mixture in a quadrupole magnetic trap (QMT), in which both species are in their fully stretched magnetic substate, i.\,e.\,\mHeB~in the $J$=1, \mjp~state and \Rb~in the \sss~state. The reason for choosing this \textit{doubly spin-stretched mixture} is that interspecies Penning and associative ionization processes (which we both will refer to as PI), i.\,e.\,
\begin{equation}
{\rm He}^*+{\rm Rb} \rightarrow \Biggl\{ \begin{array}{lr}{\rm He}+{\rm Rb}^++{\rm e}^- \\ {\rm HeRb}^++{\rm e}^-  \end{array}\label{PI}
\end{equation}
are expected to be suppressed because of spin-conservation \cite{vassen2012cat}. An upper limit for the loss rate coefficient of 5$\times$10$^{-12}$~cm$^3$s$^{-1}$ at 0.2~mK has been experimentally obtained by measuring the ion production rate \cite{byron2010sop}. For other spin-mixtures large loss rate coefficients on the order of 10$^{-10}$~cm$^3$s$^{-1}$ are expected \cite{vassen2012cat}. Simultaneous laser cooling and trapping of \mHeB~and \Rb~has already been demonstrated by the Truscott group \cite{byron2010tli,byron2010sop}.

\subsection{Theoretical description of interspecies thermalization in QMT}\label{thermalizationtheory}

Interspecies thermalization of ultracold mixtures has been described in detail in many papers (see e.\,g.\,\cite{mosk2001scw,*mudrich2002scw,*silber2005qdm,*marzok2007uto,*tassy2010sci,*ivanov2011sci}), although mostly for a Ioffe-Pritchard type of magnetic trap or optical dipole traps, i.\,e.\, a harmonic trapping potential. Here we consider thermalization for a QMT, i.\,e.\, a linear trapping potential, which requires the inclusion of Majorana heating.

The time evolution of the temperature difference $T_1-T_2$ in a two-species mixture is described by
\begin{equation}\label{thermalization}
\frac{d}{dt}\left(T_1-T_2\right)=-\gamma_{\rm th}\left(T_1-T_2\right),
\end{equation}
with thermalization rate $\gamma_{\rm th}=\gamma_{\rm coll}\xi/2.7$, where $\gamma_{\rm coll}$ is the collision rate and $\xi=4m_1m_2/(m_1+m_2)^2$. For equal mass systems 2.7 collisions are required for thermalization \cite{wu1996dso}, which can be generalized to 2.7/$\xi$ for non-equal masses \cite{mosk2001scw}. The collision rate is given by $\gamma_{\rm coll}=\sigma\langle v\rangle\langle n\rangle$, with the interspecies elastic cross section $\sigma$, the mean velocity $\langle v\rangle=\sqrt{8k_B/\pi\left(T_1/m_1+T_2/m_2\right)}$ and the mean density
$\langle n\rangle=\left(1/N_1+1/N_2\right)\int n_1(\vec{r})n_2(\vec{r}) d\vec{r}$. The temperature dependence of $\sigma$ will be discussed in Appendix~\ref{app3}.

In a QMT the density distribution (assuming an infinitely deep trap) is given by
\begin{equation}\label{equationdensityMT}
n(x,y,z)=n_0 \exp\left[-\frac{\mu \alpha \sqrt{x^2+4y^2+z^2}-m g z}{k_B T}\right],
\end{equation}
where in our case the axial direction of the coils ($y$-axis) is in the horizontal plane, $\mu$ is the magnetic moment, $\alpha$ is the magnetic field gradient along the weak (radial) axis, $g$ is the gravitational acceleration and the peak density
\begin{equation}
n_0=\frac{N}{4\pi}\left(\frac{\mu \alpha}{k_B T}\right)^3\left[1-\left(\frac{mg}{\mu\alpha}\right)^2\right]^2.
\end{equation}
In our case we can safely neglect the effect of gravity, as $mg/\mu\alpha$ is small (0.13 for \Rb, 0.003 for \mHeB), and the reduction of the overlap $\langle n\rangle$ caused by the gravitational sag of the \Rb~distribution is less than 3\%. In this approximation, 
$\int n_1(\vec{r})n_2(\vec{r})d\vec{r}=\left(\alpha^3 N_1N_2/4\pi k_B^3\right)\left(T_1/\mu_1+T_2/\mu_2\right)^{-3}$ and the thermalization rate is given by
\begin{equation}
\gamma_{\rm th} =\frac{\sigma\xi\alpha^3(N_1+N_2)}{2.7\sqrt{2}\pi^{3/2}k_B^{5/2}}\frac{\sqrt{\frac{T_1}{m_1}+\frac{T_2}{m_2}}}{\left(\frac{T_1}{\mu_1}+\frac{T_2}{\mu_2}\right)^3}.\label{thermanalytic}
\end{equation}
In case $N_1\gg N_2$, $\gamma_{\rm th}$ does not depend on $N_2$. 

In addition we have to include the Majorana effect, i.\,e.\,nonadiabatic spin flips to untrapped states at the magnetic field zero at the center of the QMT, which leads to both losses and heating, and therefore limits evaporative cooling \cite{petrich1995stc,*davis1995eco}, but also interspecies thermalization and sympathetic cooling. The Majorana heating rate is described by \cite{dubessy2012rbe,*dubessy2013erratum}
\begin{equation}\label{Majoranaheating}
\frac{d}{dt}T=\frac{\gamma_{\rm Maj}}{2 T},
\end{equation}
where $\gamma_{\rm Maj}=(8/9)\chi(\hbar/m)(2\mu\alpha/k_B)^2$, and $\chi$ is a dimensionless factor. The solution of Eq.~\ref{Majoranaheating} is given by $T(t)=\sqrt{T_0^2+\gamma_{\rm Maj} t}$, where $T_0$ is the initial temperature. 

\begin{figure}
\includegraphics[width=8.5cm]{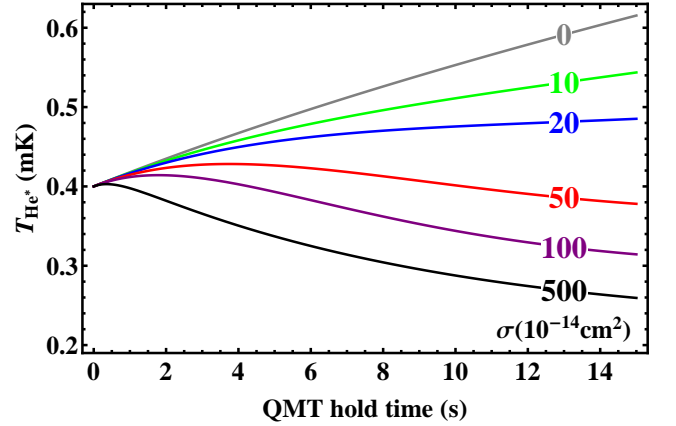}
\caption{(Color online) Calculated thermalization curves for different interspecies elastic cross sections $\sigma$, with $\alpha=120$~G/cm, $\chi$=0.1, initial temperature $T_{\rm He^*}=0.4$~mK, and atom numbers $N_{\rm He^*}^{0}=5\times10^6$ and $N_{\rm Rb}^{0}=2\times10^8$, where for \mHeB~and \Rb~we have included a one-body loss rate of (30~s)$^{-1}$ and (15~s)$^{-1}$, respectively. For the \Rb~temperature we assume $T_{\rm Rb}=(T_0-T_f)e^{-\beta t}+T_f$, with $T_0$=0.4~mK, $T_f$=0.2~mK and $\beta=0.1$~s$^{-1}$. 
\label{calculatedthermalizationcurves}}
\end{figure}

The combined effect of interspecies thermalization and Majorana heating is then described by
\begin{eqnarray}\label{temperaturediffeqgeneral}
\frac{d}{dt}\left(T_1-T_2\right)&=&-\gamma_{\rm th}\left(T_1-T_2\right)+\frac{\gamma_{\rm Maj, 1}}{2 T_1}-\frac{\gamma_{\rm Maj, 2}}{2 T_2}\\
&&+\left[\frac{d T_1}{dt}\right]_{\rm ev}-\left[\frac{d T_2}{dt}\right]_{\rm ev},\nonumber
\end{eqnarray}
where the last two terms include the effect of evaporative cooling. For our experimental parameters we can neglect Majorana heating for \Rb~($\gamma_{\rm Maj}$ is a factor 87 smaller than that of \mHeB). As we will show below, the time evolution of the \Rb~temperature is due to plain evaporation, i.\,e.\, $d T_{\rm Rb}/dt=\left[d T_{\rm Rb}/dt\right]_{\rm ev}$, whereas for \mHeB~the trap depth is too large for evaporative cooling, i.\,e.\, $\left[d T_{\rm He^*}/dt\right]_{\rm ev}=0$. This all means that for our situation we can effectively simplify Eq.~\ref{temperaturediffeqgeneral} to:
\begin{equation}\label{temperaturediffeq}
\frac{d}{dt}T_{\rm He^*}=-\gamma_{\rm th}\left(T_{\rm He^*}-T_{\rm Rb}\right)+\frac{\gamma_{\rm Maj, He^*}}{2 T_{\rm He^*}},
\end{equation}
where it is important to note that $\gamma_{\rm th}$ depends on $T_{\rm He^*}$, $T_{\rm Rb}$ and $N_{\rm Rb}$, which all change during the hold time in the QMT. The solution of Eq.~\ref{temperaturediffeq} for different values of $\sigma$ is shown in Fig.~\ref{calculatedthermalizationcurves}.

\begin{figure*}
\includegraphics[width=16cm]{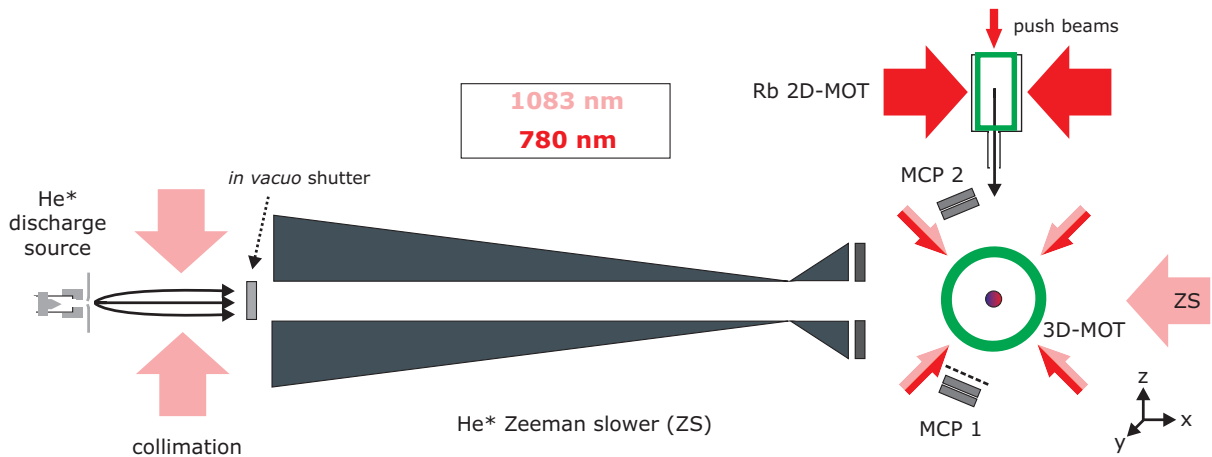}
\caption{(Color online) A schematic overview of the experimental setup and laser beams. The vertical direction is along the $z$-axis. Note that the pair of laser beams along the $y$-axis of the collimation section, the 2D-MOT and the 3D-MOT are not shown. Optical pumping and absorption imaging beams are along the $y$-axis (ZS: Zeeman slower; MCP: micro-channel plate). 
\label{setup}}
\end{figure*}

\subsection{Experimental setup}\label{experimentalsetup}

A schematic of the setup is shown in Fig.~\ref{setup}. We have added a 2D-MOT for Rb on one of the viewports of the stainless steel vacuum chamber of an existing \mHe~setup \cite{stas2004smo}. We use a liquid-nitrogen cooled dc-discharge source to produce a \HeB~beam with a \mHeB~fraction of 10$^{-4}$ \cite{stas2006hic}. The \mHeB~beam is collimated with a total power of about 500~mW, slowed in a 2.5~m long Zeeman slower and loaded into the 3D-MOT. An \textit{in vacuo} shutter is opened only during the loading time of the \mHeB~3D-MOT. Without the shutter the lifetime of \mHeB~and \Rb~atoms in the QMT is limited to less than two seconds. The pressure in the main vacuum chamber is $1\times10^{-10}$~mbar.

For the detection of \mHeB~two micro-channel plate (MCP) detectors are placed at a distance of 106~mm from the trap center, both under an angle of 22$^\circ$ with respect to the direction of gravity. MCP 1 is behind a grounded grid and detects \mHeB~atoms after release from the trap, resulting in a time-of-flight (TOF) signal that contains information about the atom number and the temperature. MCP 2 is not shielded and therefore also collects all ions (He$^+$, Rb$^+$) produced via PI. We have calibrated the MCP signals using the saturated fluorescence method, collecting transient fluorescence from a retroreflected high-power resonant beam \cite{dall2007bec}. For \Rb~we use standard absorption imaging (along $y$-direction) to obtain the atom number and temperature.

One pair of water-cooled coils provides the magnetic field gradient for both the 3D-MOT and QMT, which has a gradient along the weak axis $\alpha=\partial B/\partial x=\partial B/\partial z=\frac{1}{2}\partial B/\partial y$ of 0.6~(G/cm)/A. The axial direction of the coils ($y$) is in the horizontal plane. The laser beams for optical pumping and absorption imaging are along the $y$-direction.

\subsection{Two species magneto-optical trap}\label{magneto-optical trap}

The three retroreflected 1-inch laser beams of the 3D-MOT are derived from single mode optical fibers, in which both wavelengths for laser cooling of \mHeB~and \Rb, 1083~nm and 780~nm, respectively, are coupled together using dichroic mirrors. In this way, the 3D-MOT laser beams of the two species are automatically overlapped. To create the proper circular polarization for both wavelengths, zero-order quarter wave plates at 920~nm are used. For \Rb~a detuning of $-15$~MHz with respect to the $F=2\rightarrow F'=3$ transition is used, and a total power of the three laser beams of $\sim$~40~mW. For \mHeB~we use a large detuning of $-32$~MHz (corresponding to 20 linewidths) to reduce the light-assisted intraspecies PI loss in the 3D-MOT, and a total laser beam power of $\sim$~30~mW. The magnetic field gradient $\alpha$ is 12~G/cm. For \Rb~an additional repumper beam on the $F=1\rightarrow F'=2$ transition is added. For \mHeB~no repumper is needed because of the absence of hyperfine structure.

\begin{figure}
\includegraphics[width=8.5cm]{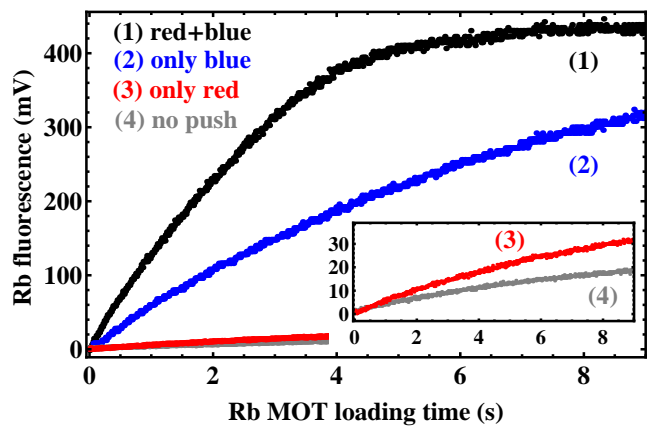}
\caption{(Color online) \Rb~fluorescence during the loading of the \Rb~3D-MOT, showing different configurations of the 2D-MOT: (1) both red- and blue-detuned push beams (black), (2) only blue-detuned push beam (blue), (3) only red-detuned push beam (red), and (4) no push beams (gray). A fluorescence signal of 400~mV corresponds to about $1\times 10^9$ atoms.
\label{RbMOTloading}}
\end{figure}

\Rb~is loaded from a 2D-MOT with a two-color push beam scheme, in which a red-detuned push beam pushes the atoms that leave the 2D-MOT in the wrong direction back towards the 3D-MOT, while a blue-detuned push beam guides the atoms through the differential pumping tube \cite{park2012cab}. The two retroreflected cooling beams are circular with a diameter of two inches and a total power  of the two laser beams of $\sim$~100~mW. The detuning is $-8.4$~MHz with respect to the $F=2\rightarrow F'=3$ transition, while the red- and the blue-detuned push beams have a detuning of $-8.4$~MHz and $+17$~MHz, respectively. In one of the cooling beams repumper light is mixed in. The differential pumping tube between the 2D- and 3D-MOT sections has a diameter of 2.5~mm and a length of 50~mm, which provides a differential pressure of $1.1\times 10^4$ between the 2D- and 3D-MOT sections. With a typical loading rate of $3\times 10^8$ atoms/s we reach $1\times10^9$ \Rb~atoms in five seconds. Fluorescence signals of the \Rb~3D-MOT loading for different configurations of the 2D-MOT are shown in Fig.~\ref{RbMOTloading}. 

\begin{figure}
\includegraphics[width=8.5cm]{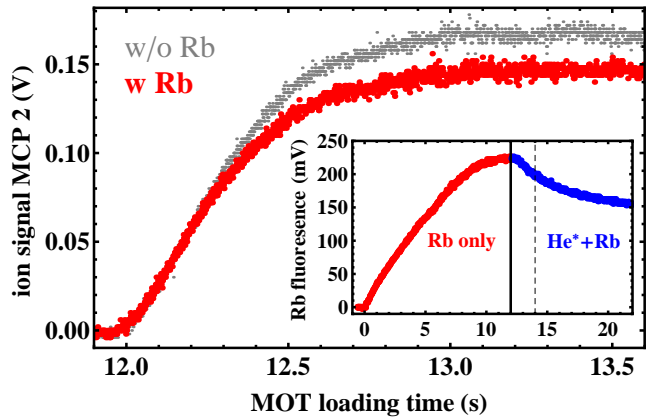}
\caption{(Color online) 
\label{HeMOTloading} Two species MOT loading. Ion signal of MCP 2 during the loading of the \mHeB~MOT, with and without the presence of the \Rb~MOT. The inset shows the \Rb~fluorescence, where after 12~s we start the \mHeB~MOT loading. The drop in \Rb~signal is mainly due to the flux of ground state He atoms. A fluorescence signal of 250~mV corresponds to about $1\times 10^9$ atoms, while the \Rb~loading rate is smaller than in Fig.~\ref{RbMOTloading}.}
\end{figure}

\mHeB~is loaded from a zero-crossing (spin-flip) Zeeman slower. The detuning of the Zeeman slowing beam is $-415$~MHz and the power is 9~mW/cm$^2$. We obtain \mHeB~3D-MOT loading rates of about $3\times10^8$ atoms/s, however, because of strong losses this results in $\sim$~$5\times10^7$ \mHeB~atoms within one second. Ion signals during the \mHeB~MOT loading are shown in Fig.~\ref{HeMOTloading}, with and without the presence of a \Rb~MOT, indicating a small decrease in \mHeB~final atom number for the two-species MOT compared to single species conditions. 

In the experimental sequence we first load \Rb~in the 3D-MOT, while only in the last two seconds we open the \textit{in vacuo} shutter to load \mHeB. During the \mHeB~loading we observe a small decrease of the \Rb~atom number (see inset of Fig.~\ref{HeMOTloading}), which is mostly due to the flux of ground state He atoms (i.\,e.\, this loss is independent on whether the Zeeman slower or \mHeB~3D-MOT light is on or not).

Afterwards, we compress the 3D-MOT by increasing the gradient to $\alpha=24$~G/cm in 70~ms, during which we increase the detunings for \Rb~and \mHeB~to $-21$~MHz and $-44$~MHz, respectively. Then we switch off the magnetic field gradient and apply optical molasses on both species for 7~ms. During this optical molasses we ramp the \Rb~detuning from $-21$~MHz to $-29$~MHz and lower the intensity of the repumper beam, while for \mHeB~we immediately jump to a detuning of $-3.5$~MHz. Finally we spin-polarize in 0.5~ms \Rb~to the \sss~state by optical pumping on the $F=2 \rightarrow F'=2$ transition with circular polarized light at a small magnetic field, while at the same time we spin-polarize \mHeB~in the \mjp~state by optical pumping on the $J=1 \rightarrow J'=2$ transition with circular polarized light. 

\begin{figure}
\includegraphics[width=8.5cm]{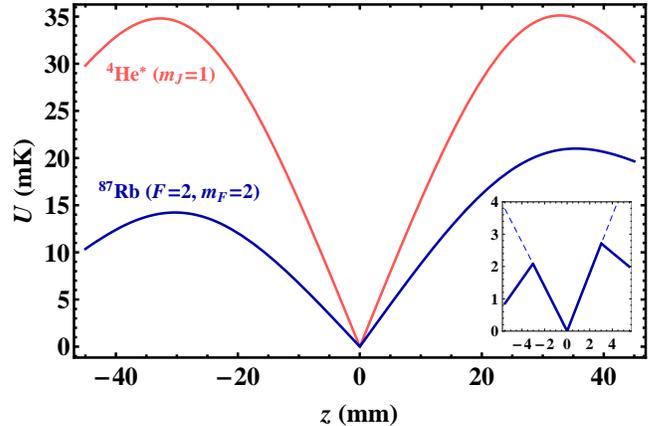}
\caption{(Color online) Trapping potentials of \mHeB~(light red) and \Rb~(dark blue) in the quadrupole magnetic trap (QMT) along the $z$-direction (i.\,e.\,along the direction of gravity), for $\alpha=120$~G/cm. Inset shows the effective potential (solid line) for \Rb~when introducing microwave (MW) radiation 75~MHz above the hyperfine splitting. The noticeable asymmetry in \Rb~potential is due to gravity. 
\label{MT_potential}}
\end{figure}

\subsection{Ultracold mixture in quadrupole magnetic trap}\label{quadrupole magnetic trap}

After the spin-polarizing pulse we ramp the magnetic field gradient within a few ms to $\alpha=48$~G/cm. After waiting for 100~ms, we ramp to $\alpha=120$~G/cm in 100~ms. The final trapping potentials are depicted in Fig.~\ref{MT_potential}. More than 95\% of the \Rb~atoms are in the \sss~state, while the \mjp~state of \mHeB~is the only magnetically trappable state. The initial atom numbers for \Rb~and \mHeB~are about $2\times10^8$ and $5\times10^6$, respectively, and their initial temperatures are both about 0.4~mK.

We hold the mixture for a variable time in the magnetic trap, after which we measure the properties of the remaining atoms by absorption imaging (\Rb) and MCP detection (\mHeB). We obtain the time evolution of the atom numbers and temperatures of the two species, in the mixture and under single species conditions. We do not observe any significant effect of \mHeB~on \Rb, both regarding atom number and temperature, which is mostly explained by the condition $N_{\rm Rb}\gg N_{\rm He^*}$. 

We apply evaporative cooling on \Rb~by shining in microwave (MW) radiation at 6910~MHz, which is 75~MHz above the hyperfine splitting, leading to an effective trap depth of 2.1~mK (see inset Fig.~\ref{MT_potential}). The time evolution of the \Rb~atom number in the QMT is shown in Fig.~\ref{RbatomMW}, with and without MW, while the \Rb~temperature is shown in Fig.~\ref{HeMTtemp} (with MW). Without MW we observe an exponential decay of the atom number with a lifetime of 36(2)~s, which is due to background collisions. With MW a stronger, non-exponential decay is visible. At a hold time of 4~s we observe from our absorption images that the \Rb~cloud becomes cross-dimensional thermalized, at which point it can be described by a single temperature of 0.33~mK. Afterwards, plain evaporation further reduce the \Rb~temperature to 0.25~mK at 14~s. 

\begin{figure}
\includegraphics[width=8.5cm]{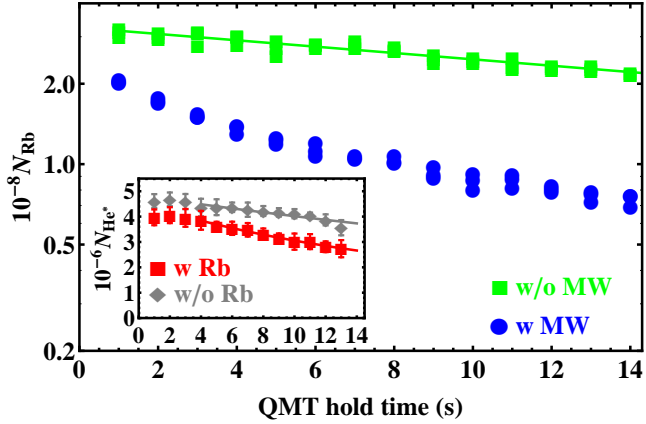}
\caption{(Color online) The number of \Rb~atoms as function of hold time in the quadrupole magnetic trap (QMT), with and without microwave (MW). The inset shows the \mHeB~atom number, with (red squares) and without (gray circles) \Rb~and the solid lines are fits of Eq.~\ref{PIloss}.\label{RbatomMW}}
\end{figure}

The time-evolutions of \mHeB~atom number and temperature are shown in Fig.~\ref{RbatomMW} and Fig.~\ref{HeMTtemp}, respectively, with and without \Rb. The initial temperature of \mHeB~is 0.40~mK, after which it increases due to Majorana heating to about 0.50~mK (with \Rb) or 0.55~mK (without \Rb) after 13~s. Thus, we observe interspecies thermalization, which however only partly counteracts Majorana heating. In Sec.~\ref{thermalizationmeasurements} we will determine the interspecies scattering length from this data. For \mHeB~the trap depth is about 30~mK (see Fig.~\ref{MT_potential}), which excludes evaporative cooling and losses.

We also observe a small reduction in the lifetime of \mHeB~in the presence of \Rb~(inset of Fig.~\ref{RbatomMW}). The time evolution of the \mHeB~atom number can be described by
\begin{equation}\label{PIloss}
\frac{d}{dt}N_{\rm He^*}=-\Gamma N_{\rm He^*}-L_2 \int n_{\rm He^*}(\vec{r})n_{\rm Rb}(\vec{r})d\vec{r},
\end{equation}
where $\Gamma$ is the one-body loss rate due to background collisions and Majorana spin-flips, and $L_2$ is the total interspecies two-body loss rate coefficient, which includes both interspecies PI and spin-relaxation. Intraspecies two-body loss, for which the loss rate coefficient is $2\times10^{-14}$~cm$^3$s$^{-1}$ \cite{borbely2012mfd}, can be fully neglected.
 
To extract $L_2$, we first fit the data without \Rb~to obtain $\Gamma$, after which we fit the data with \Rb~to obtain $L_2$. We only give an upper limit of $L_2$ because the observed reduction in lifetime may also be explained by a few percent of \Rb~atoms in the $F$=2, $m_F$=1 or $F$=1, $m_F$=$-1$ states, for which PI is not suppressed, or an increase in the Majorana spin-flip loss rate because of the smaller temperature in the presence of \Rb. We find $L_2^{\rm upper}=1.5\times 10^{-12}$~cm$^3$s$^{-1}$, which includes the estimated 50\%~systematic uncertainty in $N_{\rm He^*}$. $L_2^{\rm upper}$ is three times lower than the reported upper limit of interspecies PI at a temperature of 0.2~mK \cite{byron2010sop}.

\begin{figure}
\includegraphics[width=8.5cm]{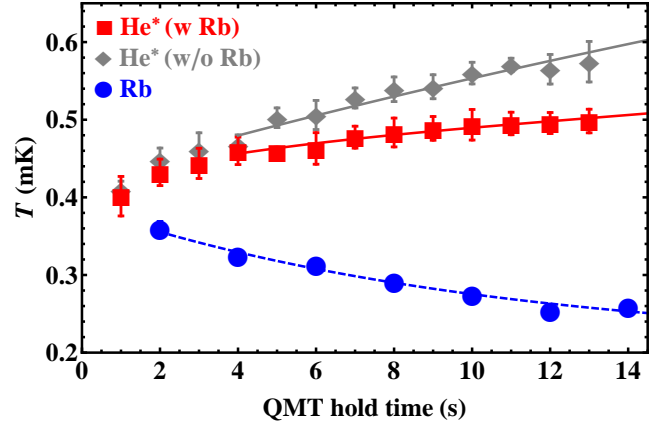}
\caption{(Color online) \mHeB~and \Rb~temperatures as function of hold time in the quadrupole magnetic trap (QMT). For \mHeB~the temperature data with \Rb~(red squares) and without \Rb~(gray triangles) are shown, together with the fits of Eq.~\ref{temperaturediffeq} (red and gray solid lines, respectively), for which only the data from 4~s on is considered. The data points are an average over four experimental runs and the error bars represent the standard deviation. The \Rb~temperature data (blue circles) are fitted by an exponential function of the form $T(t)=(T_0-T_f) e^{-\beta t}+T_f$ (blue dashed line), which is used as input for solving Eq.~\ref{temperaturediffeq}. \label{HeMTtemp}}
\end{figure}

\subsection{Determination of the interspecies scattering length from thermalization measurements}\label{thermalizationmeasurements}

To extract the interspecies elastic cross section $\sigma$ from our data that is displayed in Fig.~\ref{HeMTtemp}, we first fit the \mHeB~temperature data without \Rb, which is only described by Majorana heating (Eq.~\ref{Majoranaheating}), and we find $\chi=0.09(1)$, similar to 0.14 for $^{23}$Na~\cite{heo2011fpo} and 0.16 for \Rb~\cite{dubessy2012rbe,*dubessy2013erratum}. Then we fit the full solution of Eq.~\ref{temperaturediffeq} to the \mHeB~temperature data with \Rb, from which we obtain $\sigma_{\rm exp}=14_{-4}^{+6}\times 10^{-14}$~cm$^2$. In this analysis we fully take into account the measured time evolution of $N_{\rm He^*}$, $N_{\rm Rb}$ and $T_{\rm Rb}$, and propagate their uncertainties (one standard deviation) to obtain the uncertainty in $\sigma_{\rm exp}$. We only fit the data for hold times from 4~s on, at which the \Rb~cloud has become cross-dimensional thermalized. Note that the intraspecies thermalization rate for \Rb~is about (0.5~s)$^{-1}$ during the whole time evolution, whereas for \mHeB~it decreases from (0.5~s)$^{-1}$ to (1.5~s)$^{-1}$.

\begin{figure}
\includegraphics[width=8.5cm]{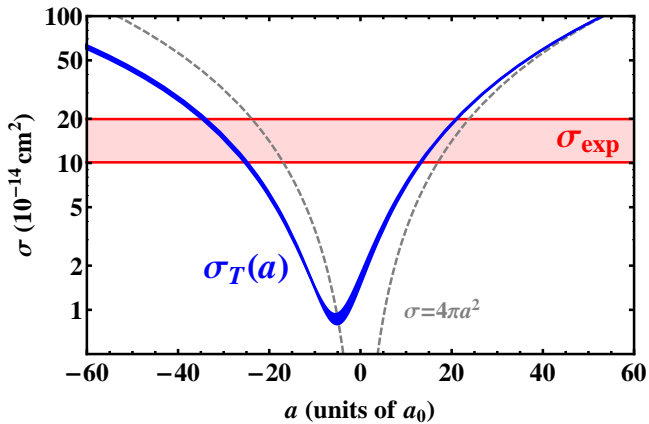}
\caption{(Color online) The experimental interspecies elastic cross section $\sigma_{\rm exp}$ (red horizontal band) for \mHeB+\Rb~and the calculated elastic cross section $\sigma_T(a)$, including the contribution from $p$-wave collisions, for $T$=0.45$\rightarrow$0.50~mK (blue shaded area). The interspecies scattering length is given by the intersections of $\sigma_{\rm exp}$ and $\sigma_T(a)$, resulting in either $a_{\rm exp}^-=-29_{-5}^{+5}$~$a_0$ or $a_{\rm exp}^+=+17_{-4}^{+4}$~$a_0$. Also the zero-temperature result $\sigma=4\pi a^2$ is shown (dashed gray line).
\label{scatteringlengthdependentcrosssection}}
\end{figure}

To relate the temperature dependent elastic cross section to the scattering length we have numerically solved the Schr\"{o}dinger equation  (see Appendix \ref{app3}). It is important to note that such a calculation is only sensitive to the long-range part of the potential, and completely independent of the short-range part obtained from the \textit{ab initio} calculations. The result is depicted in Fig.~\ref{scatteringlengthdependentcrosssection}, showing the elastic cross section $\sigma_T(a)$ for the relevant temperature range of 0.45~mK to 0.50~mK as function of scattering length (blue shaded area), which clearly deviates from the zero-temperature limit $\sigma=4\pi a^2$ (gray dashed line). With our experimental value of $\sigma_{\rm exp}$ (red horizontal band), we find the scattering length to be either $a_{\rm exp}^-=-29_{-5}^{+5}$~$a_0$ or $a_{\rm exp}^+=+17_{-4}^{+4}$~$a_0$. Note that for the doubly spin-stretched mixture, scattering only occurs in the \quartet~potential, and the experimentally obtained scattering length is the pure quartet scattering length.

\section{Results and Conclusions}\label{resultsconclusion}

The theoretically and experimentally obtained quartet scattering lengths are compared in Table~\ref{scatteringlengthtable}. One recognizes that $a_{\rm exp}^+$ is in excellent agreement with the \textit{ab initio} calculations, whereas $a_{\rm exp}^-$ can be fully excluded. In fact, with the bounds of the \textit{ab initio} calculations, we can conclude that $a_{4+87}=+17^{+1}_{-4}$~$a_0$.

In conclusion, we have performed \textit{ab initio} calculations of the \quartet~potential for \mHe+Rb, from which we have obtained the  interspecies scattering lengths for all four isotope combinations of doubly spin-stretched \mHe+Rb systems. We have determined experimentally the interspecies elastic cross section for \mHeB+\Rb~from thermalization measurements. Our combined theoretical and experimental work provides tight bounds on the interspecies scattering length, which prior to this work was completely unknown. In addition, from our experimental data we obtain an upper limit of the total interspecies two-body loss rate coefficient of $L_2^{\rm upper}=1.5\times 10^{-12}$~cm$^3$s$^{-1}$, which is three times lower than the previous reported upper limit for interspecies PI.  

The success of the \textit{ab initio} calculations, being able to quantitatively predict the scattering length for a system containing a heavy, many-electron atom, is linked to the small reduced mass and shallow \quartet~potential of the \mHe+Rb system. This leads to a small number of bound states, which reduces the sensitivity of the scattering length to the potential energy curve. Still, to achieve an 1\% accuracy of the \textit{ab initio} calculation is a formidable task for a many-electron system. We expect the same level of accuracy for the \quartet~potentials of any other combination of \mHe~with an alkali-metal atom. 

\section{Outlook}\label{outlook}

The newly obtained knowledge on the scattering lengths is crucial for realizing and exploring quantum degenerate \mHe+Rb mixtures. For example, the small interspecies scattering lengths will hamper sympathetic cooling of \mHe~by Rb, and either RF-induced forced evaporation cooling of \mHeB~or sympathetic cooling of \mHeF~with a third species, for which \mHeB~would be an excellent choice \cite{mcnamara2006dgb}, is required. Also, on basis of the intra- and interspecies scattering lengths we expect the dual BEC of \mHeB+\Rb~to be miscible and stable \cite{esry1997hft,*law1997ssi}. 

The applicability of the ultracold \mHe+Rb mixture to universal few-body physics, such as the investigation of the Efimov trimer spectrum, crucially depends on the availability and characteristics of interspecies Feshbach resonances. For this purpose close-coupling calculations that include the \doublet~potential are required. However, \textit{ab initio} calculations of the \doublet~potential are expected to be less accurate than those for the \quartet~potential, because the \doublet~potential is much deeper \cite{ruf1987tio} and supports many more bound states. Experimentally, thermalization measurements in different spin-mixtures might reveal information about the doublet scattering length, however, because Penning ionization is not suppressed, these measurements will be limited by a short lifetime. Therefore we propose to experimentally search for narrow interspecies Feshbach resonances induced by the spin-spin interaction for a mixture prepared in the lower doubly spin-stretched state, which requires a mixture in an optical dipole trap. The positions of these resonances would reveal the binding energy of the least-bound doublet level, which would provide sufficient information about the \doublet~potential.

\begin{acknowledgments}

The Amsterdam group acknowledges Jacques Bouma and Rob Kortekaas for technical support, as well as the mechanical and electronic workshops, in particular Mario Molenaar, Jurgen Buske and Niels Althuisius. We thank Tim van Leent for experimental work on part of the \mHeB~optical setup, and Rob van Rooij and Joe Borbely for their help on the computer control of the experiment. We acknowledge Servaas Kokkelmans for helpful discussions on the elastic cross section calculations. This work was financially supported by the Netherlands Organization for Scientific Research (NWO) via a VIDI grant (680-47-511) and the Dutch Foundation for Fundamental Research on Matter (FOM) via a Projectruimte grant (11PR2905). P.\,S.\,\.{Z}.\, is grateful for the support of the Foundation for Polish Science Homing Plus Programme no.\,2011-3/14 cofinanced by the European Regional Development Fund. D.\,K.\, acknowledges support from NCN grant DEC-2012/07/B/ST4/01347.

\end{acknowledgments}

\appendix 

\section{CCSD(T) calculations}

\subsection{Extrapolation towards small internuclear distances}\label{app1}

Because the CCSD(T) equations are divergent for the internuclear distances smaller than $r=8$~$a_0$, for which the interaction energy is approximately $-200$~cm$^{-1}$, we have to extrapolate our results for $r\geq 8$~$a_0$ towards smaller $r$ in order to describe the repulsive wall up to positive interaction energies. To justify the extrapolation procedure, we have calculated the Hartree-Fock interaction energy ($E_{\rm HF}$) and correlation contribution to the interaction energy obtained from the coupled cluster doubles (CCD) method ($E_{\rm CCD}$) at distances close to the inner turning point at about $r=7.6$~$a_0$. The results are shown in Fig.~\ref{ccd}. Both the Hartree-Fock and correlation contributions behave exponentially \cite{tang1984ais}, which allows extrapolation to distances at which the interaction energy becomes positive. 

\begin{figure}
\includegraphics[width=8.5cm]{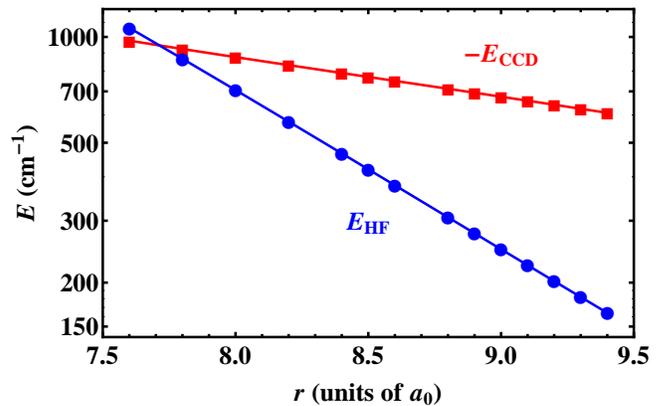}
\caption{ The contributions of Hartree-Fock ($E_{\rm HF}$) (blue circles) and correlation effects to the total interaction energy in the CCD method ($E_{\rm CCD}$) (red squares), both showing an exponential behavior (note the log scale on the Y-axis; lines are a guide to the eye). The two contributions have opposite sign and cancel each other at the inner turning point of the potential energy curve.}
\label{ccd}
\end{figure}

\subsection{Estimate of accuracy CCSD(T) method}\label{app2}

To test the error beyond the CCSD(T) method expansion, we have performed coupled cluster calculations of the interaction energy with singly-, doubly-, triply- and quadruply excited amplitudes (CCSDT and CCSDTQ, respectively) for $r=9.4$~$a_0$, using the approach introduced by K{\'a}llay \cite{kallay2001hei}. Since the cost of performing these calculations is many orders of magnitude higher compared to CCSD(T), we have to restrict ourselves to an 11 valence electrons effective core potential (ECP) and a basis set limited to $spd$ orbitals, respectively, and investigated the \textit{difference} with respect to the CCSD(T) interaction energy obtained, \textit{within} the ECP method. 

We have found that the inclusion of the full set of triple excitations leads to an increase of the well depth $D_e$ by approximately 3.7~cm$^{-1}$ compared to the CCSD(T) calculations ($D_e$=398.3~cm$^{-1}$ and 394.6~cm$^{-1}$, respectively). By further reduction of the basis set (to $sp$ orbitals) we have also found that taking into account quadruple excitations has the opposite effect: $D_e$ decreases by 0.3~cm$^{-1}$ compared to the CCSDT calculations. Hence, we can expect that the systematic error due to the incompleteness of the CCSD(T) method should be limited by the difference between CCSDT and CCSD(T) well depths, which in recommended basis set should be proportional to the $D_e$ ratio in the limited basis sets. This leads to a systematic error of +4.3 cm$^{-1}$.

\section{Determination of the interspecies scattering length from elastic cross section}\label{app3}

\begin{figure}
\includegraphics[width=8.5cm]{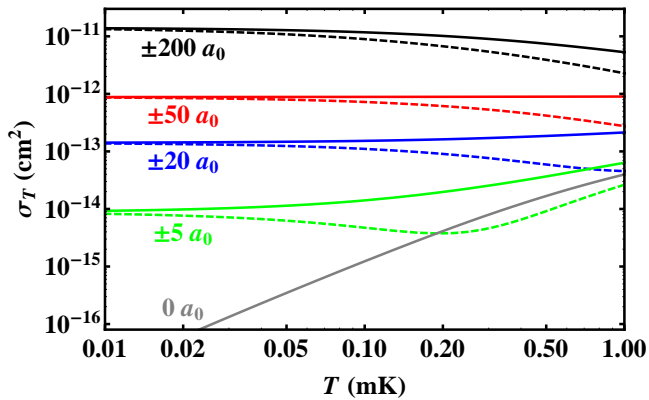}
\caption{(Color online) Numerical results of $\sigma_T$ for $a$=$\pm$5~$a_0$ (green), $\pm$20~$a_0$ (blue), $\pm$50~$a_0$ (red), $\pm$200~$a_0$ (black), where the solid (dashed) lines are representing the positive (negative) values of $a$, and $a=0$~$a_0$ (gray).
\label{temperaturedependentcrosssection}}
\end{figure}

In the zero-temperature limit the elastic cross section $\sigma$ is simply related to the $s$-wave scattering length $a$ via $\sigma=4\pi a^2$. However, in the temperature range of our measurement, we do not fulfill this limit. Therefore, we have performed numerical calculations on basis of the radial Schr\"{o}dinger equation to obtain the connection between $a$ and $\sigma$ at the experimentally relevant temperature range. Here we use a simple Lennard-Jones potential to demonstrate that for this particular purpose only knowledge of the long-range potential is sufficient, and the experimental scattering length determination is completely independent of the \textit{ab initio} calculations of the short-range potential. The Lennard-Jones potential is given by $V_{b}(r)=-(C_6/r^6)\left[1-b/r^6\right]$, where we can tune the depth of the potential, and therefore $a$, by parameter $b$. Note that the less accurate $C_6$ coefficient of Zhang \etal~\cite{zhang2012lri} gives the same result than the one obtained in this work, showing that a few percent accuracy is sufficient.

The energy dependent elastic cross section is given by $\sigma_E=\sum_{l=0}^\infty\sigma_E^l$, where $\sigma_E^l=(4\pi/k^2)(2l+1)\sin\delta_l(k)^2$, $\delta_l$ is the $l$-wave phase shift, $k=\sqrt{2\mu_r E}/\hbar$, $\mu_r$ is the reduced mass, and $E$ is the collision energy. The scattering length is defined by $a=-\lim\limits_{k\rightarrow 0}\tan\delta_0(k)/k$. The temperature dependent partial cross section $\sigma_T^l$ is obtained by taking the Boltzmann average over $\sigma_E^l$:
\begin{equation}
\sigma_T^l=\frac{1}{(k_B T)^{2}}\int_0^\infty \sigma_E^lE e^{-E/k_B T}dE
\end{equation}
where one has to consider an effective temperature, given by $T=\mu_r (T_{\rm He^*}/m_{\rm He^*}+T_{\rm Rb}/m_{\rm Rb})\approx T_{\rm He^*}$. The total temperature dependent cross section is $\sigma_T=\sum_{l=0}^\infty\sigma_T^l$. We find that $\sigma_E^l$ (and therefore also $\sigma_T$) is only dependent on $b$ via $a$, but different values of $b$ that give the same $a$, also give the same $\sigma_E^l$. This means that our results are independent on the particular choice of model potential, and that $\sigma_T$ is fully determined by $a$ and $C_6$ (and the reduced mass $\mu_r$).

In Fig.~\ref{temperaturedependentcrosssection} we show $\sigma_T$ for several values of $a$, where the solid (dashed) lines represent the positive (negative) values of $a$. Because the $p$-wave centrifugal barrier height is 3.4~mK, in most cases $\sigma_T\approx\sigma_T^0$ for $T\leq 1$~mK. However, the contribution of $p$-wave collisions, $\sigma_T^1$, can still be significant for small values of $|a|$, for which $\sigma_T^0$ itself is very small. Therefore for all calculations we include $p$-wave collisions, i.\,e.\, $\sigma_T=\sigma_T^0+\sigma_T^1$. In the temperature range of 0.1$-$1~mK deviations from the zero-temperature cross sections are significant, especially for small $|a|$. In general, for a given $|a|$, the cross section for $a>0$ is larger than $a<0$. Finally, we calculate $\sigma_T$ as a function of $a$ for the experimental relevant temperature range, for which the result is shown in Fig.~\ref{scatteringlengthdependentcrosssection}.

\bibliography{HeRbLib}

\begin{thebibliography}{52}%
\makeatletter
\providecommand \@ifxundefined [1]{%
 \@ifx{#1\undefined}
}%
\providecommand \@ifnum [1]{%
 \ifnum #1\expandafter \@firstoftwo
 \else \expandafter \@secondoftwo
 \fi
}%
\providecommand \@ifx [1]{%
 \ifx #1\expandafter \@firstoftwo
 \else \expandafter \@secondoftwo
 \fi
}%
\providecommand \natexlab [1]{#1}%
\providecommand \enquote  [1]{``#1''}%
\providecommand \bibnamefont  [1]{#1}%
\providecommand \bibfnamefont [1]{#1}%
\providecommand \citenamefont [1]{#1}%
\providecommand \href@noop [0]{\@secondoftwo}%
\providecommand \href [0]{\begingroup \@sanitize@url \@href}%
\providecommand \@href[1]{\@@startlink{#1}\@@href}%
\providecommand \@@href[1]{\endgroup#1\@@endlink}%
\providecommand \@sanitize@url [0]{\catcode `\\12\catcode `\$12\catcode
  `\&12\catcode `\#12\catcode `\^12\catcode `\_12\catcode `\%12\relax}%
\providecommand \@@startlink[1]{}%
\providecommand \@@endlink[0]{}%
\providecommand \url  [0]{\begingroup\@sanitize@url \@url }%
\providecommand \@url [1]{\endgroup\@href {#1}{\urlprefix }}%
\providecommand \urlprefix  [0]{URL }%
\providecommand \Eprint [0]{\href }%
\providecommand \doibase [0]{http://dx.doi.org/}%
\providecommand \selectlanguage [0]{\@gobble}%
\providecommand \bibinfo  [0]{\@secondoftwo}%
\providecommand \bibfield  [0]{\@secondoftwo}%
\providecommand \translation [1]{[#1]}%
\providecommand \BibitemOpen [0]{}%
\providecommand \bibitemStop [0]{}%
\providecommand \bibitemNoStop [0]{.\EOS\space}%
\providecommand \EOS [0]{\spacefactor3000\relax}%
\providecommand \BibitemShut  [1]{\csname bibitem#1\endcsname}%
\let\auto@bib@innerbib\@empty
\bibitem [{\citenamefont {Carr}\ \emph {et~al.}(2009)\citenamefont {Carr},
  \citenamefont {DeMille}, \citenamefont {Krems},\ and\ \citenamefont
  {Ye}}]{carr2009cau}%
  \BibitemOpen
  \bibfield  {author} {\bibinfo {author} {\bibfnamefont {L.~D.}\ \bibnamefont
  {Carr}}, \bibinfo {author} {\bibfnamefont {D.}~\bibnamefont {DeMille}},
  \bibinfo {author} {\bibfnamefont {R.~V.}\ \bibnamefont {Krems}}, \ and\
  \bibinfo {author} {\bibfnamefont {J.}~\bibnamefont {Ye}},\ }\href@noop {}
  {\bibfield  {journal} {\bibinfo  {journal} {New J.\ Phys.}\ }\textbf
  {\bibinfo {volume} {11}},\ \bibinfo {pages} {055049} (\bibinfo {year}
  {2009})}\BibitemShut {NoStop}%
\bibitem [{\citenamefont {Levinsen}\ \emph {et~al.}(2009)\citenamefont
  {Levinsen}, \citenamefont {Tiecke}, \citenamefont {Walraven},\ and\
  \citenamefont {Petrov}}]{levinsen2009ads}%
  \BibitemOpen
  \bibfield  {author} {\bibinfo {author} {\bibfnamefont {J.}~\bibnamefont
  {Levinsen}}, \bibinfo {author} {\bibfnamefont {T.~G.}\ \bibnamefont
  {Tiecke}}, \bibinfo {author} {\bibfnamefont {J.~T.~M.}\ \bibnamefont
  {Walraven}}, \ and\ \bibinfo {author} {\bibfnamefont {D.~S.}\ \bibnamefont
  {Petrov}},\ }\href@noop {} {\bibfield  {journal} {\bibinfo  {journal} {Phys.\
  Rev.\ Lett.}\ }\textbf {\bibinfo {volume} {103}},\ \bibinfo {pages} {153202}
  (\bibinfo {year} {2009})}\BibitemShut {NoStop}%
\bibitem [{\citenamefont {Castin}\ \emph {et~al.}(2010)\citenamefont {Castin},
  \citenamefont {Mora},\ and\ \citenamefont {Pricoupenko}}]{castin2010fbe}%
  \BibitemOpen
  \bibfield  {author} {\bibinfo {author} {\bibfnamefont {Y.}~\bibnamefont
  {Castin}}, \bibinfo {author} {\bibfnamefont {C.}~\bibnamefont {Mora}}, \ and\
  \bibinfo {author} {\bibfnamefont {L.}~\bibnamefont {Pricoupenko}},\
  }\href@noop {} {\bibfield  {journal} {\bibinfo  {journal} {Phys.\ Rev.\
  Lett.}\ }\textbf {\bibinfo {volume} {105}},\ \bibinfo {pages} {223201}
  (\bibinfo {year} {2010})}\BibitemShut {NoStop}%
\bibitem [{\citenamefont {Mathy}\ \emph {et~al.}(2011)\citenamefont {Mathy},
  \citenamefont {Parish},\ and\ \citenamefont {Huse}}]{mathy2011tma}%
  \BibitemOpen
  \bibfield  {author} {\bibinfo {author} {\bibfnamefont {C.~J.~M.}\
  \bibnamefont {Mathy}}, \bibinfo {author} {\bibfnamefont {M.~M.}\ \bibnamefont
  {Parish}}, \ and\ \bibinfo {author} {\bibfnamefont {D.~A.}\ \bibnamefont
  {Huse}},\ }\href@noop {} {\bibfield  {journal} {\bibinfo  {journal} {Phys.\
  Rev.\ Lett.}\ }\textbf {\bibinfo {volume} {106}},\ \bibinfo {pages} {166404}
  (\bibinfo {year} {2011})}\BibitemShut {NoStop}%
\bibitem [{\citenamefont {Blume}(2012)}]{blume2012ufb}%
  \BibitemOpen
  \bibfield  {author} {\bibinfo {author} {\bibfnamefont {D.}~\bibnamefont
  {Blume}},\ }\href@noop {} {\bibfield  {journal} {\bibinfo  {journal} {Phys.\
  Rev.\ Lett.}\ }\textbf {\bibinfo {volume} {109}},\ \bibinfo {pages} {230404}
  (\bibinfo {year} {2012})}\BibitemShut {NoStop}%
\bibitem [{\citenamefont {Efremov}\ \emph {et~al.}(2013)\citenamefont
  {Efremov}, \citenamefont {Plimak}, \citenamefont {Ivanov},\ and\
  \citenamefont {Schleich}}]{efremov2013tbb}%
  \BibitemOpen
  \bibfield  {author} {\bibinfo {author} {\bibfnamefont {M.~A.}\ \bibnamefont
  {Efremov}}, \bibinfo {author} {\bibfnamefont {L.}~\bibnamefont {Plimak}},
  \bibinfo {author} {\bibfnamefont {M.~Y.}\ \bibnamefont {Ivanov}}, \ and\
  \bibinfo {author} {\bibfnamefont {W.~P.}\ \bibnamefont {Schleich}},\
  }\href@noop {} {\bibfield  {journal} {\bibinfo  {journal} {Phys.\ Rev.\
  Lett.}\ }\textbf {\bibinfo {volume} {111}},\ \bibinfo {pages} {113201}
  (\bibinfo {year} {2013})}\BibitemShut {NoStop}%
\bibitem [{\citenamefont {Zinner}(2013)}]{zinner2013eso}%
  \BibitemOpen
  \bibfield  {author} {\bibinfo {author} {\bibfnamefont {N.~T.}\ \bibnamefont
  {Zinner}},\ }\href@noop {} {\bibfield  {journal} {\bibinfo  {journal}
  {Europhys.\ Lett.}\ }\textbf {\bibinfo {volume} {101}},\ \bibinfo {pages}
  {60009} (\bibinfo {year} {2013})}\BibitemShut {NoStop}%
\bibitem [{\citenamefont {Hu}\ \emph {et~al.}()\citenamefont {Hu},
  \citenamefont {Ma\'{s}ka}, \citenamefont {Clark},\ and\ \citenamefont
  {Freericks}}]{hu2014oog}%
  \BibitemOpen
  \bibfield  {author} {\bibinfo {author} {\bibfnamefont {A.}~\bibnamefont
  {Hu}}, \bibinfo {author} {\bibfnamefont {M.~M.}\ \bibnamefont {Ma\'{s}ka}},
  \bibinfo {author} {\bibfnamefont {C.~W.}\ \bibnamefont {Clark}}, \ and\
  \bibinfo {author} {\bibfnamefont {J.~K.}\ \bibnamefont {Freericks}},\
  }\href@noop {} {}\Eprint {http://arxiv.org/abs/1407.1000} {arXiv:1407.1000}
  \BibitemShut {NoStop}%
\bibitem [{\citenamefont {Pires}\ \emph {et~al.}(2014)\citenamefont {Pires},
  \citenamefont {Ulmanis}, \citenamefont {H\"afner}, \citenamefont {Repp},
  \citenamefont {Arias}, \citenamefont {Kuhnle},\ and\ \citenamefont
  {Weidem\"uller}}]{pires2014ooe}%
  \BibitemOpen
  \bibfield  {author} {\bibinfo {author} {\bibfnamefont {R.}~\bibnamefont
  {Pires}}, \bibinfo {author} {\bibfnamefont {J.}~\bibnamefont {Ulmanis}},
  \bibinfo {author} {\bibfnamefont {S.}~\bibnamefont {H\"afner}}, \bibinfo
  {author} {\bibfnamefont {M.}~\bibnamefont {Repp}}, \bibinfo {author}
  {\bibfnamefont {A.}~\bibnamefont {Arias}}, \bibinfo {author} {\bibfnamefont
  {E.~D.}\ \bibnamefont {Kuhnle}}, \ and\ \bibinfo {author} {\bibfnamefont
  {M.}~\bibnamefont {Weidem\"uller}},\ }\href@noop {} {\bibfield  {journal}
  {\bibinfo  {journal} {Phys.\ Rev.\ Lett.}\ }\textbf {\bibinfo {volume}
  {112}},\ \bibinfo {pages} {250404} (\bibinfo {year} {2014})}\BibitemShut
  {NoStop}%
\bibitem [{\citenamefont {Tung}\ \emph {et~al.}()\citenamefont {Tung},
  \citenamefont {Jim\'{e}nez-Garc\'{\i}a}, \citenamefont {Johansen},
  \citenamefont {Parker},\ and\ \citenamefont {Chin}}]{tung2014oog}%
  \BibitemOpen
  \bibfield  {author} {\bibinfo {author} {\bibfnamefont {S.-K.}\ \bibnamefont
  {Tung}}, \bibinfo {author} {\bibfnamefont {K.}~\bibnamefont
  {Jim\'{e}nez-Garc\'{\i}a}}, \bibinfo {author} {\bibfnamefont
  {J.}~\bibnamefont {Johansen}}, \bibinfo {author} {\bibfnamefont
  {C.}~\bibnamefont {Parker}}, \ and\ \bibinfo {author} {\bibfnamefont
  {C.}~\bibnamefont {Chin}},\ }\href@noop {} {}\Eprint
  {http://arxiv.org/abs/1402.5943} {arXiv:1402.5943} \BibitemShut {NoStop}%
\bibitem [{\citenamefont {Nemitz}\ \emph {et~al.}(2009)\citenamefont {Nemitz},
  \citenamefont {Baumer}, \citenamefont {M\"{u}nchow}, \citenamefont {Tassy},\
  and\ \citenamefont {G\"{o}rlitz}}]{nemitz2009poh}%
  \BibitemOpen
  \bibfield  {author} {\bibinfo {author} {\bibfnamefont {N.}~\bibnamefont
  {Nemitz}}, \bibinfo {author} {\bibfnamefont {F.}~\bibnamefont {Baumer}},
  \bibinfo {author} {\bibfnamefont {F.}~\bibnamefont {M\"{u}nchow}}, \bibinfo
  {author} {\bibfnamefont {S.}~\bibnamefont {Tassy}}, \ and\ \bibinfo {author}
  {\bibfnamefont {A.}~\bibnamefont {G\"{o}rlitz}},\ }\href@noop {} {\bibfield
  {journal} {\bibinfo  {journal} {Phys.\ Rev.\ A}\ }\textbf {\bibinfo {volume}
  {79}},\ \bibinfo {pages} {061403(R)} (\bibinfo {year} {2009})}\BibitemShut
  {NoStop}%
\bibitem [{\citenamefont {Hara}\ \emph {et~al.}(2011)\citenamefont {Hara},
  \citenamefont {Takasu}, \citenamefont {Yamaoka}, \citenamefont {Doyle},\ and\
  \citenamefont {Takahashi}}]{hara2011qdm}%
  \BibitemOpen
  \bibfield  {author} {\bibinfo {author} {\bibfnamefont {H.}~\bibnamefont
  {Hara}}, \bibinfo {author} {\bibfnamefont {Y.}~\bibnamefont {Takasu}},
  \bibinfo {author} {\bibfnamefont {Y.}~\bibnamefont {Yamaoka}}, \bibinfo
  {author} {\bibfnamefont {J.~M.}\ \bibnamefont {Doyle}}, \ and\ \bibinfo
  {author} {\bibfnamefont {Y.}~\bibnamefont {Takahashi}},\ }\href@noop {}
  {\bibfield  {journal} {\bibinfo  {journal} {Phys.\ Rev.\ Lett.}\ }\textbf
  {\bibinfo {volume} {106}},\ \bibinfo {pages} {205304} (\bibinfo {year}
  {2011})}\BibitemShut {NoStop}%
\bibitem [{\citenamefont {Hansen}\ \emph {et~al.}(2011)\citenamefont {Hansen},
  \citenamefont {Khramov}, \citenamefont {Dowd}, \citenamefont {Jamison},
  \citenamefont {Ivanov},\ and\ \citenamefont {Gupta}}]{hansen2011qdm}%
  \BibitemOpen
  \bibfield  {author} {\bibinfo {author} {\bibfnamefont {A.~H.}\ \bibnamefont
  {Hansen}}, \bibinfo {author} {\bibfnamefont {A.}~\bibnamefont {Khramov}},
  \bibinfo {author} {\bibfnamefont {W.~H.}\ \bibnamefont {Dowd}}, \bibinfo
  {author} {\bibfnamefont {A.~O.}\ \bibnamefont {Jamison}}, \bibinfo {author}
  {\bibfnamefont {V.~V.}\ \bibnamefont {Ivanov}}, \ and\ \bibinfo {author}
  {\bibfnamefont {S.}~\bibnamefont {Gupta}},\ }\href@noop {} {\bibfield
  {journal} {\bibinfo  {journal} {Phys.\ Rev.\ A}\ }\textbf {\bibinfo {volume}
  {84}},\ \bibinfo {pages} {011606(R)} (\bibinfo {year} {2011})}\BibitemShut
  {NoStop}%
\bibitem [{\citenamefont {Pasquiou}\ \emph {et~al.}(2013)\citenamefont
  {Pasquiou}, \citenamefont {Bayerle}, \citenamefont {Tzanova}, \citenamefont
  {Stellmer}, \citenamefont {Szczepkowski}, \citenamefont {Parigger},
  \citenamefont {Grimm},\ and\ \citenamefont {Schreck}}]{pasquiou2013qdm}%
  \BibitemOpen
  \bibfield  {author} {\bibinfo {author} {\bibfnamefont {B.}~\bibnamefont
  {Pasquiou}}, \bibinfo {author} {\bibfnamefont {A.}~\bibnamefont {Bayerle}},
  \bibinfo {author} {\bibfnamefont {S.~M.}\ \bibnamefont {Tzanova}}, \bibinfo
  {author} {\bibfnamefont {S.}~\bibnamefont {Stellmer}}, \bibinfo {author}
  {\bibfnamefont {J.}~\bibnamefont {Szczepkowski}}, \bibinfo {author}
  {\bibfnamefont {M.}~\bibnamefont {Parigger}}, \bibinfo {author}
  {\bibfnamefont {R.}~\bibnamefont {Grimm}}, \ and\ \bibinfo {author}
  {\bibfnamefont {F.}~\bibnamefont {Schreck}},\ }\href@noop {} {\bibfield
  {journal} {\bibinfo  {journal} {Phys.\ Rev.\ A}\ }\textbf {\bibinfo {volume}
  {88}},\ \bibinfo {pages} {023601} (\bibinfo {year} {2013})}\BibitemShut
  {NoStop}%
\bibitem [{\citenamefont {Vassen}\ \emph {et~al.}(2012)\citenamefont {Vassen},
  \citenamefont {Cohen-Tannoudji}, \citenamefont {Leduc}, \citenamefont
  {Boiron}, \citenamefont {Westbrook}, \citenamefont {Truscott}, \citenamefont
  {Baldwin}, \citenamefont {Birkl}, \citenamefont {Cancio},\ and\ \citenamefont
  {Trippenbach}}]{vassen2012cat}%
  \BibitemOpen
  \bibfield  {author} {\bibinfo {author} {\bibfnamefont {W.}~\bibnamefont
  {Vassen}}, \bibinfo {author} {\bibfnamefont {C.}~\bibnamefont
  {Cohen-Tannoudji}}, \bibinfo {author} {\bibfnamefont {M.}~\bibnamefont
  {Leduc}}, \bibinfo {author} {\bibfnamefont {D.}~\bibnamefont {Boiron}},
  \bibinfo {author} {\bibfnamefont {C.}~\bibnamefont {Westbrook}}, \bibinfo
  {author} {\bibfnamefont {A.}~\bibnamefont {Truscott}}, \bibinfo {author}
  {\bibfnamefont {K.}~\bibnamefont {Baldwin}}, \bibinfo {author} {\bibfnamefont
  {G.}~\bibnamefont {Birkl}}, \bibinfo {author} {\bibfnamefont
  {P.}~\bibnamefont {Cancio}}, \ and\ \bibinfo {author} {\bibfnamefont
  {M.}~\bibnamefont {Trippenbach}},\ }\href@noop {} {\bibfield  {journal}
  {\bibinfo  {journal} {Rev.\ Mod.\ Phys.}\ }\textbf {\bibinfo {volume} {84}},\
  \bibinfo {pages} {175} (\bibinfo {year} {2012})}\BibitemShut {NoStop}%
\bibitem [{\citenamefont {Byron}\ \emph
  {et~al.}(2010{\natexlab{a}})\citenamefont {Byron}, \citenamefont {Dall},
  \citenamefont {Rugway},\ and\ \citenamefont {Truscott}}]{byron2010sop}%
  \BibitemOpen
  \bibfield  {author} {\bibinfo {author} {\bibfnamefont {L.~J.}\ \bibnamefont
  {Byron}}, \bibinfo {author} {\bibfnamefont {R.~G.}\ \bibnamefont {Dall}},
  \bibinfo {author} {\bibfnamefont {W.}~\bibnamefont {Rugway}}, \ and\ \bibinfo
  {author} {\bibfnamefont {A.~G.}\ \bibnamefont {Truscott}},\ }\href@noop {}
  {\bibfield  {journal} {\bibinfo  {journal} {New. J. Phys.}\ }\textbf
  {\bibinfo {volume} {12}},\ \bibinfo {pages} {013004} (\bibinfo {year}
  {2010}{\natexlab{a}})}\BibitemShut {NoStop}%
\bibitem [{\citenamefont {Przybytek}\ and\ \citenamefont
  {Jeziorski}(2005)}]{przybytek2005bft}%
  \BibitemOpen
  \bibfield  {author} {\bibinfo {author} {\bibfnamefont {M.}~\bibnamefont
  {Przybytek}}\ and\ \bibinfo {author} {\bibfnamefont {B.}~\bibnamefont
  {Jeziorski}},\ }\href@noop {} {\bibfield  {journal} {\bibinfo  {journal} {J.\
  Chem.\ Phys.}\ }\textbf {\bibinfo {volume} {123}},\ \bibinfo {pages} {134315}
  (\bibinfo {year} {2005})}\BibitemShut {NoStop}%
\bibitem [{\citenamefont {Knowles}\ \emph {et~al.}(1993)\citenamefont
  {Knowles}, \citenamefont {Hampel},\ and\ \citenamefont
  {Werner}}]{knowles1993cct}%
  \BibitemOpen
  \bibfield  {author} {\bibinfo {author} {\bibfnamefont {P.~J.}\ \bibnamefont
  {Knowles}}, \bibinfo {author} {\bibfnamefont {C.}~\bibnamefont {Hampel}}, \
  and\ \bibinfo {author} {\bibfnamefont {H.~J.}\ \bibnamefont {Werner}},\
  }\href@noop {} {\bibfield  {journal} {\bibinfo  {journal} {J.\ Chem.\ Phys.}\
  }\textbf {\bibinfo {volume} {99}},\ \bibinfo {pages} {5219} (\bibinfo {year}
  {1993})}\BibitemShut {NoStop}%
\bibitem [{\citenamefont {Werner}\ \emph {et~al.}(2012)\citenamefont {Werner},
  \citenamefont {Knowles}, \citenamefont {Knizia}, \citenamefont {Manby},
  \citenamefont {{Sch\"{u}tz}} \emph {et~al.}}]{MOLPRO_brief}%
  \BibitemOpen
  \bibfield  {author} {\bibinfo {author} {\bibfnamefont {H.-J.}\ \bibnamefont
  {Werner}}, \bibinfo {author} {\bibfnamefont {P.~J.}\ \bibnamefont {Knowles}},
  \bibinfo {author} {\bibfnamefont {G.}~\bibnamefont {Knizia}}, \bibinfo
  {author} {\bibfnamefont {F.~R.}\ \bibnamefont {Manby}}, \bibinfo {author}
  {\bibfnamefont {M.}~\bibnamefont {{Sch\"{u}tz}}},  \emph {et~al.},\
  }\href@noop {} {\enquote {\bibinfo {title} {{MOLPRO}, version 2012.1, a
  package of \textit{ab initio} programs},}\ } (\bibinfo {year} {2012}),\
  \bibinfo {note} {see http://www.molpro.net}\BibitemShut {NoStop}%
\bibitem [{\citenamefont {Reiher}\ and\ \citenamefont
  {Wolf}(2004)}]{reiher2004edo}%
  \BibitemOpen
  \bibfield  {author} {\bibinfo {author} {\bibfnamefont {M.}~\bibnamefont
  {Reiher}}\ and\ \bibinfo {author} {\bibfnamefont {A.}~\bibnamefont {Wolf}},\
  }\href@noop {} {\bibfield  {journal} {\bibinfo  {journal} {J.\ Chem.\ Phys.}\
  }\textbf {\bibinfo {volume} {121}},\ \bibinfo {pages} {2037} (\bibinfo {year}
  {2004})}\BibitemShut {NoStop}%
\bibitem [{\citenamefont {Hapka}\ \emph {et~al.}(2013)\citenamefont {Hapka},
  \citenamefont {Cha{\l}asiski}, \citenamefont {K{\l}os},\ and\ \citenamefont
  {\.{Z}uchowski}}]{hapka2013fpi}%
  \BibitemOpen
  \bibfield  {author} {\bibinfo {author} {\bibfnamefont {M.}~\bibnamefont
  {Hapka}}, \bibinfo {author} {\bibfnamefont {G.}~\bibnamefont
  {Cha{\l}asiski}}, \bibinfo {author} {\bibfnamefont {J.}~\bibnamefont
  {K{\l}os}}, \ and\ \bibinfo {author} {\bibfnamefont {P.~S.}\ \bibnamefont
  {\.{Z}uchowski}},\ }\href@noop {} {\bibfield  {journal} {\bibinfo  {journal}
  {J. Chem. Phys.}\ }\textbf {\bibinfo {volume} {139}},\ \bibinfo {pages}
  {014307} (\bibinfo {year} {2013})}\BibitemShut {NoStop}%
\bibitem [{\citenamefont {Roos}\ \emph {et~al.}(2003)\citenamefont {Roos},
  \citenamefont {Veryazov},\ and\ \citenamefont {Widmark}}]{roos2003ran}%
  \BibitemOpen
  \bibfield  {author} {\bibinfo {author} {\bibfnamefont {B.~O.}\ \bibnamefont
  {Roos}}, \bibinfo {author} {\bibfnamefont {V.}~\bibnamefont {Veryazov}}, \
  and\ \bibinfo {author} {\bibfnamefont {P.-O.}\ \bibnamefont {Widmark}},\
  }\href@noop {} {\bibfield  {journal} {\bibinfo  {journal} {Theor.\ Chem.\
  Acc.}\ }\textbf {\bibinfo {volume} {111}},\ \bibinfo {pages} {345} (\bibinfo
  {year} {2003})}\BibitemShut {NoStop}%
\bibitem [{\citenamefont {Boys}\ and\ \citenamefont
  {Bernardi}(1970)}]{boys1970tco}%
  \BibitemOpen
  \bibfield  {author} {\bibinfo {author} {\bibfnamefont {S.~F.}\ \bibnamefont
  {Boys}}\ and\ \bibinfo {author} {\bibfnamefont {F.}~\bibnamefont
  {Bernardi}},\ }\href@noop {} {\bibfield  {journal} {\bibinfo  {journal}
  {Mol.\ Phys.}\ }\textbf {\bibinfo {volume} {19}},\ \bibinfo {pages} {553}
  (\bibinfo {year} {1970})}\BibitemShut {NoStop}%
\bibitem [{\citenamefont {Tang}\ and\ \citenamefont
  {Toennies}(1984)}]{tang1984ais}%
  \BibitemOpen
  \bibfield  {author} {\bibinfo {author} {\bibfnamefont {K.~T.}\ \bibnamefont
  {Tang}}\ and\ \bibinfo {author} {\bibfnamefont {J.~P.}\ \bibnamefont
  {Toennies}},\ }\href@noop {} {\bibfield  {journal} {\bibinfo  {journal} {J.\
  Chem.\ Phys.}\ }\textbf {\bibinfo {volume} {80}},\ \bibinfo {pages} {3726}
  (\bibinfo {year} {1984})}\BibitemShut {NoStop}%
\bibitem [{\citenamefont {Zhang}\ \emph {et~al.}(2012)\citenamefont {Zhang},
  \citenamefont {Tang}, \citenamefont {Shi}, \citenamefont {Yan},\ and\
  \citenamefont {Schwingenschl\"{o}gl}}]{zhang2012lri}%
  \BibitemOpen
  \bibfield  {author} {\bibinfo {author} {\bibfnamefont {J.-Y.}\ \bibnamefont
  {Zhang}}, \bibinfo {author} {\bibfnamefont {L.-Y.}\ \bibnamefont {Tang}},
  \bibinfo {author} {\bibfnamefont {T.-Y.}\ \bibnamefont {Shi}}, \bibinfo
  {author} {\bibfnamefont {Z.-C.}\ \bibnamefont {Yan}}, \ and\ \bibinfo
  {author} {\bibfnamefont {U.}~\bibnamefont {Schwingenschl\"{o}gl}},\
  }\href@noop {} {\bibfield  {journal} {\bibinfo  {journal} {Phys.\ Rev.\ A}\
  }\textbf {\bibinfo {volume} {86}},\ \bibinfo {pages} {064701} (\bibinfo
  {year} {2012})}\BibitemShut {NoStop}%
\bibitem [{\citenamefont {Yan}\ and\ \citenamefont {Babb}(1998)}]{yan1998lri}%
  \BibitemOpen
  \bibfield  {author} {\bibinfo {author} {\bibfnamefont {Z.-C.}\ \bibnamefont
  {Yan}}\ and\ \bibinfo {author} {\bibfnamefont {J.~F.}\ \bibnamefont {Babb}},\
  }\href@noop {} {\bibfield  {journal} {\bibinfo  {journal} {Phys.\ Rev.\ A}\
  }\textbf {\bibinfo {volume} {58}},\ \bibinfo {pages} {1247} (\bibinfo {year}
  {1998})}\BibitemShut {NoStop}%
\bibitem [{\citenamefont {Porsev}\ \emph {et~al.}(2014)\citenamefont {Porsev},
  \citenamefont {Safronova}, \citenamefont {Derevianko},\ and\ \citenamefont
  {Clark}}]{porsev2014rmb}%
  \BibitemOpen
  \bibfield  {author} {\bibinfo {author} {\bibfnamefont {S.~G.}\ \bibnamefont
  {Porsev}}, \bibinfo {author} {\bibfnamefont {M.~S.}\ \bibnamefont
  {Safronova}}, \bibinfo {author} {\bibfnamefont {A.}~\bibnamefont
  {Derevianko}}, \ and\ \bibinfo {author} {\bibfnamefont {C.~W.}\ \bibnamefont
  {Clark}},\ }\href@noop {} {\bibfield  {journal} {\bibinfo  {journal} {Phys.\
  Rev.\ A}\ }\textbf {\bibinfo {volume} {89}},\ \bibinfo {pages} {022703}
  (\bibinfo {year} {2014})}\BibitemShut {NoStop}%
\bibitem [{\citenamefont {LeRoy}\ \emph {et~al.}(2006)\citenamefont {LeRoy},
  \citenamefont {Huang},\ and\ \citenamefont {Jary}}]{leroy2006aaa}%
  \BibitemOpen
  \bibfield  {author} {\bibinfo {author} {\bibfnamefont {R.~J.}\ \bibnamefont
  {LeRoy}}, \bibinfo {author} {\bibfnamefont {Y.}~\bibnamefont {Huang}}, \ and\
  \bibinfo {author} {\bibfnamefont {C.}~\bibnamefont {Jary}},\ }\href@noop {}
  {\bibfield  {journal} {\bibinfo  {journal} {J.\ Chem.\ Phys.}\ }\textbf
  {\bibinfo {volume} {125}},\ \bibinfo {eid} {164310} (\bibinfo {year}
  {2006})}\BibitemShut {NoStop}%
\bibitem [{\citenamefont {K{\'a}llay}\ and\ \citenamefont
  {Surj{\'a}n}(2001)}]{kallay2001hei}%
  \BibitemOpen
  \bibfield  {author} {\bibinfo {author} {\bibfnamefont {M.}~\bibnamefont
  {K{\'a}llay}}\ and\ \bibinfo {author} {\bibfnamefont {P.~R.}\ \bibnamefont
  {Surj{\'a}n}},\ }\href@noop {} {\bibfield  {journal} {\bibinfo  {journal}
  {J.\ Chem.\ Phys.}\ }\textbf {\bibinfo {volume} {115}},\ \bibinfo {pages}
  {2945} (\bibinfo {year} {2001})}\BibitemShut {NoStop}%
\bibitem [{\citenamefont {Strauss}\ \emph {et~al.}(2010)\citenamefont
  {Strauss}, \citenamefont {Takekoshi}, \citenamefont {Lang}, \citenamefont
  {Winkler}, \citenamefont {Grimm}, \citenamefont {{Hecker Denschlag}},\ and\
  \citenamefont {Tiemann}}]{strauss2010hra}%
  \BibitemOpen
  \bibfield  {author} {\bibinfo {author} {\bibfnamefont {C.}~\bibnamefont
  {Strauss}}, \bibinfo {author} {\bibfnamefont {T.}~\bibnamefont {Takekoshi}},
  \bibinfo {author} {\bibfnamefont {F.}~\bibnamefont {Lang}}, \bibinfo {author}
  {\bibfnamefont {K.}~\bibnamefont {Winkler}}, \bibinfo {author} {\bibfnamefont
  {R.}~\bibnamefont {Grimm}}, \bibinfo {author} {\bibfnamefont
  {J.}~\bibnamefont {{Hecker Denschlag}}}, \ and\ \bibinfo {author}
  {\bibfnamefont {E.}~\bibnamefont {Tiemann}},\ }\href@noop {} {\bibfield
  {journal} {\bibinfo  {journal} {Phys.\ Rev.\ A}\ }\textbf {\bibinfo {volume}
  {82}},\ \bibinfo {pages} {052514} (\bibinfo {year} {2010})}\BibitemShut
  {NoStop}%
\bibitem [{\citenamefont {Byron}\ \emph
  {et~al.}(2010{\natexlab{b}})\citenamefont {Byron}, \citenamefont {Dall},\
  and\ \citenamefont {Truscott}}]{byron2010tli}%
  \BibitemOpen
  \bibfield  {author} {\bibinfo {author} {\bibfnamefont {L.~J.}\ \bibnamefont
  {Byron}}, \bibinfo {author} {\bibfnamefont {R.~G.}\ \bibnamefont {Dall}}, \
  and\ \bibinfo {author} {\bibfnamefont {A.~G.}\ \bibnamefont {Truscott}},\
  }\href@noop {} {\bibfield  {journal} {\bibinfo  {journal} {Phys.\ Rev.\ A}\
  }\textbf {\bibinfo {volume} {81}},\ \bibinfo {pages} {013405} (\bibinfo
  {year} {2010}{\natexlab{b}})}\BibitemShut {NoStop}%
\bibitem [{\citenamefont {Mosk}\ \emph {et~al.}(2001)\citenamefont {Mosk},
  \citenamefont {Kraft}, \citenamefont {Mudrich}, \citenamefont {Singer},
  \citenamefont {Wohlleben}, \citenamefont {Grimm},\ and\ \citenamefont
  {Weidem\"{u}ller}}]{mosk2001scw}%
  \BibitemOpen
  \bibfield  {author} {\bibinfo {author} {\bibfnamefont {A.}~\bibnamefont
  {Mosk}}, \bibinfo {author} {\bibfnamefont {S.}~\bibnamefont {Kraft}},
  \bibinfo {author} {\bibfnamefont {M.}~\bibnamefont {Mudrich}}, \bibinfo
  {author} {\bibfnamefont {K.}~\bibnamefont {Singer}}, \bibinfo {author}
  {\bibfnamefont {W.}~\bibnamefont {Wohlleben}}, \bibinfo {author}
  {\bibfnamefont {R.}~\bibnamefont {Grimm}}, \ and\ \bibinfo {author}
  {\bibfnamefont {M.}~\bibnamefont {Weidem\"{u}ller}},\ }\href@noop {}
  {\bibfield  {journal} {\bibinfo  {journal} {Appl.\ Phys.\ B}\ }\textbf
  {\bibinfo {volume} {73}},\ \bibinfo {pages} {791} (\bibinfo {year}
  {2001})}\BibitemShut {NoStop}%
\bibitem [{\citenamefont {Mudrich}\ \emph {et~al.}(2002)\citenamefont
  {Mudrich}, \citenamefont {Kraft}, \citenamefont {Singer}, \citenamefont
  {Grimm}, \citenamefont {Mosk},\ and\ \citenamefont
  {Weidem\"{u}ller}}]{mudrich2002scw}%
  \BibitemOpen
  \bibfield  {author} {\bibinfo {author} {\bibfnamefont {M.}~\bibnamefont
  {Mudrich}}, \bibinfo {author} {\bibfnamefont {S.}~\bibnamefont {Kraft}},
  \bibinfo {author} {\bibfnamefont {K.}~\bibnamefont {Singer}}, \bibinfo
  {author} {\bibfnamefont {R.}~\bibnamefont {Grimm}}, \bibinfo {author}
  {\bibfnamefont {A.}~\bibnamefont {Mosk}}, \ and\ \bibinfo {author}
  {\bibfnamefont {M.}~\bibnamefont {Weidem\"{u}ller}},\ }\href@noop {}
  {\bibfield  {journal} {\bibinfo  {journal} {Phys.\ Rev.\ Lett.}\ }\textbf
  {\bibinfo {volume} {88}},\ \bibinfo {pages} {253001} (\bibinfo {year}
  {2002})}\BibitemShut {NoStop}%
\bibitem [{\citenamefont {Silber}\ \emph {et~al.}(2005)\citenamefont {Silber},
  \citenamefont {G\"{u}nther}, \citenamefont {Marzok}, \citenamefont {Deh},
  \citenamefont {{Ph. W. Courteille}},\ and\ \citenamefont
  {Zimmermann}}]{silber2005qdm}%
  \BibitemOpen
  \bibfield  {author} {\bibinfo {author} {\bibfnamefont {C.}~\bibnamefont
  {Silber}}, \bibinfo {author} {\bibfnamefont {S.}~\bibnamefont {G\"{u}nther}},
  \bibinfo {author} {\bibfnamefont {C.}~\bibnamefont {Marzok}}, \bibinfo
  {author} {\bibfnamefont {B.}~\bibnamefont {Deh}}, \bibinfo {author}
  {\bibnamefont {{Ph. W. Courteille}}}, \ and\ \bibinfo {author} {\bibfnamefont
  {C.}~\bibnamefont {Zimmermann}},\ }\href@noop {} {\bibfield  {journal}
  {\bibinfo  {journal} {Phys.\ Rev.\ Lett.}\ }\textbf {\bibinfo {volume}
  {95}},\ \bibinfo {pages} {170408} (\bibinfo {year} {2005})}\BibitemShut
  {NoStop}%
\bibitem [{\citenamefont {Marzok}\ \emph {et~al.}(2007)\citenamefont {Marzok},
  \citenamefont {Deh}, \citenamefont {{Ph. W. Courteille}},\ and\ \citenamefont
  {Zimmermann}}]{marzok2007uto}%
  \BibitemOpen
  \bibfield  {author} {\bibinfo {author} {\bibfnamefont {C.}~\bibnamefont
  {Marzok}}, \bibinfo {author} {\bibfnamefont {B.}~\bibnamefont {Deh}},
  \bibinfo {author} {\bibnamefont {{Ph. W. Courteille}}}, \ and\ \bibinfo
  {author} {\bibfnamefont {C.}~\bibnamefont {Zimmermann}},\ }\href@noop {}
  {\bibfield  {journal} {\bibinfo  {journal} {Phys.\ Rev.\ A}\ }\textbf
  {\bibinfo {volume} {76}},\ \bibinfo {pages} {052704} (\bibinfo {year}
  {2007})}\BibitemShut {NoStop}%
\bibitem [{\citenamefont {Tassy}\ \emph {et~al.}(2010)\citenamefont {Tassy},
  \citenamefont {Nemitz}, \citenamefont {Baumer}, \citenamefont {H\"{o}hl},
  \citenamefont {Bat\"{a}r},\ and\ \citenamefont {G\"{o}rlitz}}]{tassy2010sci}%
  \BibitemOpen
  \bibfield  {author} {\bibinfo {author} {\bibfnamefont {S.}~\bibnamefont
  {Tassy}}, \bibinfo {author} {\bibfnamefont {N.}~\bibnamefont {Nemitz}},
  \bibinfo {author} {\bibfnamefont {F.}~\bibnamefont {Baumer}}, \bibinfo
  {author} {\bibfnamefont {C.}~\bibnamefont {H\"{o}hl}}, \bibinfo {author}
  {\bibfnamefont {A.}~\bibnamefont {Bat\"{a}r}}, \ and\ \bibinfo {author}
  {\bibfnamefont {A.}~\bibnamefont {G\"{o}rlitz}},\ }\href@noop {} {\bibfield
  {journal} {\bibinfo  {journal} {J.\ Phys.\ B: At.\ Mol.\ Opt.\ Phys.}\
  }\textbf {\bibinfo {volume} {43}},\ \bibinfo {pages} {205309} (\bibinfo
  {year} {2010})}\BibitemShut {NoStop}%
\bibitem [{\citenamefont {Ivanov}\ \emph {et~al.}(2011)\citenamefont {Ivanov},
  \citenamefont {Khramov}, \citenamefont {Hansen}, \citenamefont {Dowd},
  \citenamefont {M\"{u}nchow}, \citenamefont {Jamison},\ and\ \citenamefont
  {Gupta}}]{ivanov2011sci}%
  \BibitemOpen
  \bibfield  {author} {\bibinfo {author} {\bibfnamefont {V.~V.}\ \bibnamefont
  {Ivanov}}, \bibinfo {author} {\bibfnamefont {A.}~\bibnamefont {Khramov}},
  \bibinfo {author} {\bibfnamefont {A.~H.}\ \bibnamefont {Hansen}}, \bibinfo
  {author} {\bibfnamefont {W.~H.}\ \bibnamefont {Dowd}}, \bibinfo {author}
  {\bibfnamefont {F.}~\bibnamefont {M\"{u}nchow}}, \bibinfo {author}
  {\bibfnamefont {A.~O.}\ \bibnamefont {Jamison}}, \ and\ \bibinfo {author}
  {\bibfnamefont {S.}~\bibnamefont {Gupta}},\ }\href@noop {} {\bibfield
  {journal} {\bibinfo  {journal} {Phys.\ Rev.\ Lett.}\ }\textbf {\bibinfo
  {volume} {106}},\ \bibinfo {pages} {153201} (\bibinfo {year}
  {2011})}\BibitemShut {NoStop}%
\bibitem [{\citenamefont {Wu}\ and\ \citenamefont {Foot}(1996)}]{wu1996dso}%
  \BibitemOpen
  \bibfield  {author} {\bibinfo {author} {\bibfnamefont {H.}~\bibnamefont
  {Wu}}\ and\ \bibinfo {author} {\bibfnamefont {C.~J.}\ \bibnamefont {Foot}},\
  }\href@noop {} {\bibfield  {journal} {\bibinfo  {journal} {J.\ Phys.\ B}\
  }\textbf {\bibinfo {volume} {29}},\ \bibinfo {pages} {L321} (\bibinfo {year}
  {1996})}\BibitemShut {NoStop}%
\bibitem [{\citenamefont {Petrich}\ \emph {et~al.}(1995)\citenamefont
  {Petrich}, \citenamefont {Anderson}, \citenamefont {Ensher},\ and\
  \citenamefont {Cornell}}]{petrich1995stc}%
  \BibitemOpen
  \bibfield  {author} {\bibinfo {author} {\bibfnamefont {W.}~\bibnamefont
  {Petrich}}, \bibinfo {author} {\bibfnamefont {M.~H.}\ \bibnamefont
  {Anderson}}, \bibinfo {author} {\bibfnamefont {J.~R.}\ \bibnamefont
  {Ensher}}, \ and\ \bibinfo {author} {\bibfnamefont {E.~A.}\ \bibnamefont
  {Cornell}},\ }\href@noop {} {\bibfield  {journal} {\bibinfo  {journal}
  {Phys.\ Rev.\ Lett.}\ }\textbf {\bibinfo {volume} {74}},\ \bibinfo {pages}
  {3352} (\bibinfo {year} {1995})}\BibitemShut {NoStop}%
\bibitem [{\citenamefont {Davis}\ \emph {et~al.}(1995)\citenamefont {Davis},
  \citenamefont {Mewes}, \citenamefont {Joffe}, \citenamefont {Andrews},\ and\
  \citenamefont {Ketterle}}]{davis1995eco}%
  \BibitemOpen
  \bibfield  {author} {\bibinfo {author} {\bibfnamefont {K.~B.}\ \bibnamefont
  {Davis}}, \bibinfo {author} {\bibfnamefont {M.-O.}\ \bibnamefont {Mewes}},
  \bibinfo {author} {\bibfnamefont {M.~A.}\ \bibnamefont {Joffe}}, \bibinfo
  {author} {\bibfnamefont {M.~R.}\ \bibnamefont {Andrews}}, \ and\ \bibinfo
  {author} {\bibfnamefont {W.}~\bibnamefont {Ketterle}},\ }\href@noop {}
  {\bibfield  {journal} {\bibinfo  {journal} {Phys.\ Rev.\ Lett.}\ }\textbf
  {\bibinfo {volume} {74}},\ \bibinfo {pages} {5202} (\bibinfo {year}
  {1995})}\BibitemShut {NoStop}%
\bibitem [{\citenamefont {Dubessy}\ \emph {et~al.}(2012)\citenamefont
  {Dubessy}, \citenamefont {Merloti}, \citenamefont {Longchambon},
  \citenamefont {Pottie}, \citenamefont {Liennard}, \citenamefont {Perrin},
  \citenamefont {Lorent},\ and\ \citenamefont {Perrin}}]{dubessy2012rbe}%
  \BibitemOpen
  \bibfield  {author} {\bibinfo {author} {\bibfnamefont {R.}~\bibnamefont
  {Dubessy}}, \bibinfo {author} {\bibfnamefont {K.}~\bibnamefont {Merloti}},
  \bibinfo {author} {\bibfnamefont {L.}~\bibnamefont {Longchambon}}, \bibinfo
  {author} {\bibfnamefont {P.-E.}\ \bibnamefont {Pottie}}, \bibinfo {author}
  {\bibfnamefont {T.}~\bibnamefont {Liennard}}, \bibinfo {author}
  {\bibfnamefont {A.}~\bibnamefont {Perrin}}, \bibinfo {author} {\bibfnamefont
  {V.}~\bibnamefont {Lorent}}, \ and\ \bibinfo {author} {\bibfnamefont
  {H.}~\bibnamefont {Perrin}},\ }\href@noop {} {\bibfield  {journal} {\bibinfo
  {journal} {Phys.\ Rev.\ A}\ }\textbf {\bibinfo {volume} {85}},\ \bibinfo
  {pages} {013643} (\bibinfo {year} {2012})}\BibitemShut {NoStop}%
\bibitem [{\citenamefont {Dubessy}\ \emph {et~al.}(2013)\citenamefont
  {Dubessy}, \citenamefont {Merloti}, \citenamefont {Longchambon},
  \citenamefont {Pottie}, \citenamefont {Liennard}, \citenamefont {Perrin},
  \citenamefont {Lorent},\ and\ \citenamefont {Perrin}}]{dubessy2013erratum}%
  \BibitemOpen
  \bibfield  {author} {\bibinfo {author} {\bibfnamefont {R.}~\bibnamefont
  {Dubessy}}, \bibinfo {author} {\bibfnamefont {K.}~\bibnamefont {Merloti}},
  \bibinfo {author} {\bibfnamefont {L.}~\bibnamefont {Longchambon}}, \bibinfo
  {author} {\bibfnamefont {P.-E.}\ \bibnamefont {Pottie}}, \bibinfo {author}
  {\bibfnamefont {T.}~\bibnamefont {Liennard}}, \bibinfo {author}
  {\bibfnamefont {A.}~\bibnamefont {Perrin}}, \bibinfo {author} {\bibfnamefont
  {V.}~\bibnamefont {Lorent}}, \ and\ \bibinfo {author} {\bibfnamefont
  {H.}~\bibnamefont {Perrin}},\ }\href@noop {} {\bibfield  {journal} {\bibinfo
  {journal} {Phys.\ Rev.\ A}\ }\textbf {\bibinfo {volume} {87}},\ \bibinfo
  {pages} {049903(E)} (\bibinfo {year} {2013})}\BibitemShut {NoStop}%
\bibitem [{\citenamefont {Stas}\ \emph {et~al.}(2004)\citenamefont {Stas},
  \citenamefont {McNamara}, \citenamefont {Hogervorst},\ and\ \citenamefont
  {Vassen}}]{stas2004smo}%
  \BibitemOpen
  \bibfield  {author} {\bibinfo {author} {\bibfnamefont {R.~J.~W.}\
  \bibnamefont {Stas}}, \bibinfo {author} {\bibfnamefont {J.~M.}\ \bibnamefont
  {McNamara}}, \bibinfo {author} {\bibfnamefont {W.}~\bibnamefont
  {Hogervorst}}, \ and\ \bibinfo {author} {\bibfnamefont {W.}~\bibnamefont
  {Vassen}},\ }\href@noop {} {\bibfield  {journal} {\bibinfo  {journal} {Phys.\
  Rev.\ Lett.}\ }\textbf {\bibinfo {volume} {93}},\ \bibinfo {pages} {053001}
  (\bibinfo {year} {2004})}\BibitemShut {NoStop}%
\bibitem [{\citenamefont {Stas}\ \emph {et~al.}(2006)\citenamefont {Stas},
  \citenamefont {McNamara}, \citenamefont {Hogervorst},\ and\ \citenamefont
  {Vassen}}]{stas2006hic}%
  \BibitemOpen
  \bibfield  {author} {\bibinfo {author} {\bibfnamefont {R.~J.~W.}\
  \bibnamefont {Stas}}, \bibinfo {author} {\bibfnamefont {J.~M.}\ \bibnamefont
  {McNamara}}, \bibinfo {author} {\bibfnamefont {W.}~\bibnamefont
  {Hogervorst}}, \ and\ \bibinfo {author} {\bibfnamefont {W.}~\bibnamefont
  {Vassen}},\ }\href@noop {} {\bibfield  {journal} {\bibinfo  {journal} {Phys.\
  Rev.\ A}\ }\textbf {\bibinfo {volume} {73}},\ \bibinfo {pages} {032713}
  (\bibinfo {year} {2006})}\BibitemShut {NoStop}%
\bibitem [{\citenamefont {Dall}\ and\ \citenamefont
  {Truscott}(2007)}]{dall2007bec}%
  \BibitemOpen
  \bibfield  {author} {\bibinfo {author} {\bibfnamefont {R.~G.}\ \bibnamefont
  {Dall}}\ and\ \bibinfo {author} {\bibfnamefont {A.~G.}\ \bibnamefont
  {Truscott}},\ }\href@noop {} {\bibfield  {journal} {\bibinfo  {journal} {Opt.
  Commun.}\ }\textbf {\bibinfo {volume} {270}},\ \bibinfo {pages} {255}
  (\bibinfo {year} {2007})}\BibitemShut {NoStop}%
\bibitem [{\citenamefont {Park}\ \emph {et~al.}(2012)\citenamefont {Park},
  \citenamefont {Noh},\ and\ \citenamefont {Mun}}]{park2012cab}%
  \BibitemOpen
  \bibfield  {author} {\bibinfo {author} {\bibfnamefont {S.~J.}\ \bibnamefont
  {Park}}, \bibinfo {author} {\bibfnamefont {J.}~\bibnamefont {Noh}}, \ and\
  \bibinfo {author} {\bibfnamefont {J.}~\bibnamefont {Mun}},\ }\href@noop {}
  {\bibfield  {journal} {\bibinfo  {journal} {Opt.\ Commun.}\ }\textbf
  {\bibinfo {volume} {285}},\ \bibinfo {pages} {3950} (\bibinfo {year}
  {2012})}\BibitemShut {NoStop}%
\bibitem [{\citenamefont {Borbely}\ \emph {et~al.}(2012)\citenamefont
  {Borbely}, \citenamefont {{van Rooij}}, \citenamefont {Knoop},\ and\
  \citenamefont {Vassen}}]{borbely2012mfd}%
  \BibitemOpen
  \bibfield  {author} {\bibinfo {author} {\bibfnamefont {J.~S.}\ \bibnamefont
  {Borbely}}, \bibinfo {author} {\bibfnamefont {R.}~\bibnamefont {{van
  Rooij}}}, \bibinfo {author} {\bibfnamefont {S.}~\bibnamefont {Knoop}}, \ and\
  \bibinfo {author} {\bibfnamefont {W.}~\bibnamefont {Vassen}},\ }\href@noop {}
  {\bibfield  {journal} {\bibinfo  {journal} {Phys.\ Rev.\ A}\ }\textbf
  {\bibinfo {volume} {85}},\ \bibinfo {pages} {022706} (\bibinfo {year}
  {2012})}\BibitemShut {NoStop}%
\bibitem [{\citenamefont {Heo}\ \emph {et~al.}(2011)\citenamefont {Heo},
  \citenamefont {Choi},\ and\ \citenamefont {Shin}}]{heo2011fpo}%
  \BibitemOpen
  \bibfield  {author} {\bibinfo {author} {\bibfnamefont {M.-S.}\ \bibnamefont
  {Heo}}, \bibinfo {author} {\bibfnamefont {J.-Y.}\ \bibnamefont {Choi}}, \
  and\ \bibinfo {author} {\bibfnamefont {Y.-I.}\ \bibnamefont {Shin}},\
  }\href@noop {} {\bibfield  {journal} {\bibinfo  {journal} {Phys.\ Rev.\ A}\
  }\textbf {\bibinfo {volume} {83}},\ \bibinfo {pages} {013622} (\bibinfo
  {year} {2011})}\BibitemShut {NoStop}%
\bibitem [{\citenamefont {McNamara}\ \emph {et~al.}(2006)\citenamefont
  {McNamara}, \citenamefont {Jeltes}, \citenamefont {Tychkov}, \citenamefont
  {Hogervorst},\ and\ \citenamefont {Vassen}}]{mcnamara2006dgb}%
  \BibitemOpen
  \bibfield  {author} {\bibinfo {author} {\bibfnamefont {J.~M.}\ \bibnamefont
  {McNamara}}, \bibinfo {author} {\bibfnamefont {T.}~\bibnamefont {Jeltes}},
  \bibinfo {author} {\bibfnamefont {A.~S.}\ \bibnamefont {Tychkov}}, \bibinfo
  {author} {\bibfnamefont {W.}~\bibnamefont {Hogervorst}}, \ and\ \bibinfo
  {author} {\bibfnamefont {W.}~\bibnamefont {Vassen}},\ }\href@noop {}
  {\bibfield  {journal} {\bibinfo  {journal} {Phys.\ Rev.\ Lett.}\ }\textbf
  {\bibinfo {volume} {97}},\ \bibinfo {pages} {080404} (\bibinfo {year}
  {2006})}\BibitemShut {NoStop}%
\bibitem [{\citenamefont {Esry}\ \emph {et~al.}(1997)\citenamefont {Esry},
  \citenamefont {Greene}, \citenamefont {Burke},\ and\ \citenamefont
  {Bohn}}]{esry1997hft}%
  \BibitemOpen
  \bibfield  {author} {\bibinfo {author} {\bibfnamefont {B.~D.}\ \bibnamefont
  {Esry}}, \bibinfo {author} {\bibfnamefont {C.~H.}\ \bibnamefont {Greene}},
  \bibinfo {author} {\bibfnamefont {J.~P.}\ \bibnamefont {Burke}}, \ and\
  \bibinfo {author} {\bibfnamefont {J.~L.}\ \bibnamefont {Bohn}},\ }\href@noop
  {} {\bibfield  {journal} {\bibinfo  {journal} {Phys.\ Rev.\ Lett.}\ }\textbf
  {\bibinfo {volume} {78}},\ \bibinfo {pages} {3594} (\bibinfo {year}
  {1997})}\BibitemShut {NoStop}%
\bibitem [{\citenamefont {Law}\ \emph {et~al.}(1997)\citenamefont {Law},
  \citenamefont {Pu}, \citenamefont {Bigelow},\ and\ \citenamefont
  {Eberly}}]{law1997ssi}%
  \BibitemOpen
  \bibfield  {author} {\bibinfo {author} {\bibfnamefont {C.~K.}\ \bibnamefont
  {Law}}, \bibinfo {author} {\bibfnamefont {H.}~\bibnamefont {Pu}}, \bibinfo
  {author} {\bibfnamefont {N.~P.}\ \bibnamefont {Bigelow}}, \ and\ \bibinfo
  {author} {\bibfnamefont {J.~H.}\ \bibnamefont {Eberly}},\ }\href@noop {}
  {\bibfield  {journal} {\bibinfo  {journal} {Phys.\ Rev.\ Lett.}\ }\textbf
  {\bibinfo {volume} {79}},\ \bibinfo {pages} {3105} (\bibinfo {year}
  {1997})}\BibitemShut {NoStop}%
\bibitem [{\citenamefont {Ruf}\ \emph {et~al.}(1987)\citenamefont {Ruf},
  \citenamefont {Yencha},\ and\ \citenamefont {Hotop}}]{ruf1987tio}%
  \BibitemOpen
  \bibfield  {author} {\bibinfo {author} {\bibfnamefont {M.-W.}\ \bibnamefont
  {Ruf}}, \bibinfo {author} {\bibfnamefont {A.~J.}\ \bibnamefont {Yencha}}, \
  and\ \bibinfo {author} {\bibfnamefont {H.}~\bibnamefont {Hotop}},\
  }\href@noop {} {\bibfield  {journal} {\bibinfo  {journal} {Z.\ Phys.\ D}\
  }\textbf {\bibinfo {volume} {5}},\ \bibinfo {pages} {9} (\bibinfo {year}
  {1987})}\BibitemShut {NoStop}%
\end{thebibliography}%

\end{document}